\documentclass[11pt]{article}
\usepackage[a4paper, margin=2.5cm]{geometry}

\usepackage[utf8]{inputenc}

\usepackage[ngerman,english, main=english]{babel}

\usepackage{csquotes}

\usepackage{verbatim}

\usepackage{datetime2}

\usepackage{blindtext}

\usepackage{authblk}

\usepackage[
	usenames,
	dvipsnames,
	hyperref]{xcolor}

\usepackage{colortbl}

\usepackage[
	unicode=true,
	colorlinks=true,
	linkcolor=red!35!black,
	citecolor=green!35!black,
	urlcolor=magenta!35!black,
]{hyperref}

\usepackage[
	natbib,
	maxcitenames=1,
	maxbibnames=100,
	style=authoryear-comp,
	uniquename=false,
	uniquelist=false,
	giveninits,
	doi=true,
	url=true,
	isbn=false,
	eprint=true,
	backend=biber,
	useprefix=true,
	hyperref,
	sorting=ynt,
	sortcites=true,
	date=year,
]{biblatex}

\usepackage{subfiles}

\usepackage{array}

\usepackage{adjustbox}
\usepackage{xspace}
\usepackage{setspace}
\usepackage{layouts}

\usepackage[
	format=plain,
	labelfont=bf
]{caption}
\usepackage{subcaption}

\usepackage{tabularx}
\newcolumntype{Y}{>{\centering\arraybackslash}X}
\usepackage{booktabs}
\usepackage{multirow}
\usepackage{longtable}
\usepackage{sidecap}

\usepackage{enumitem}

\usepackage{graphicx}

\usepackage{placeins}

\usepackage[
	ruled,
	vlined
]{algorithm2e}

\usepackage{amsmath}
\usepackage{amssymb}
\usepackage{bm}
\usepackage[tbtags]{mathtools}
\usepackage{xfrac}
\usepackage{mleftright}
\usepackage{xifthen}

\usepackage[
	range-units=single,
	per-mode=symbol
]{siunitx}

\usepackage[
	acronyms,
	toc,
	nomain,
	nopostdot,
	nogroupskip,
	indexonlyfirst,
	shortcuts,
	sort=standard
]{glossaries}
\usepackage[nogroupskip]{glossaries-extra}

\usepackage[
	capitalize,
	noabbrev
]{cleveref}

\usepackage{subdepth}

\usepackage{catchfile}

\usepackage[all]{nowidow}

\usepackage{pifont}

\usepackage{afterpage}

\usepackage{pdflscape}

\graphicspath{{figures}}

\AtEveryBibitem{%
	\clearfield{isbn}%
	\clearfield{issn}%
	\clearfield{eprint}%
	\clearfield{urldate}%
	\clearname{editor}%
	\clearname{publisher}%
	\clearfield{place}%
	\ifentrytype{article}{
		\clearfield{url}%
		\clearfield{urlyear}%
		\clearfield{location}%
	}{}%
	\ifentrytype{inproceedings}{
		\clearfield{url}%
		\clearfield{urlyear}%
		\clearfield{location}%
	}{}%
	\ifentrytype{book}{
		\clearfield{url}%
		\clearfield{urlyear}%
		\clearfield{location}%
	}{}%
	\ifentrytype{unpublished}{
		\clearfield{urldate}%
		\clearfield{date}%
	}{}%
}

\AtEveryBibitem{%
	\ifentrytype{unpublished}{%
		\clearfield{urldate}%
	}{}%
}

\crefname{algocf}{Algorithm}{Algorithms}
\crefname{table}{Table}{Tables}
\crefname{chapter}{Chapter}{Chapters}
\crefname{equation}{Equation}{Equations}
\crefname{section}{Section}{Sections}

\creflabelformat{equation}{#2\textup{#1}#3}

\ExplSyntaxOn

\NewDocumentCommand{\getenv}{om}
 {
  \sys_get_shell:nnN { kpsewhich ~ --var-value ~ #2 } { } \l_tmpa_tl
  \tl_trim_spaces:N \l_tmpa_tl
  \IfNoValueTF { #1 }
   {
    \tl_use:N \l_tmpa_tl
   }
   {
    \tl_set_eq:NN #1 \l_tmpa_tl
   }
 }

 \tl_const:Nn \c_getenv_par_tl { \par }
 \prg_new_protected_conditional:Nnn \septatrix_if_setenv:n {T,F,TF}
 {
	 \sys_get_shell:nnN { kpsewhich ~ --var-value ~ #1 } { } \l_tmpa_tl
	 \tl_if_eq:NNTF \l_tmpa_tl \c_getenv_par_tl
	 { \prg_return_false: }
	 { \prg_return_true: }
 }

\NewDocumentCommand{\ifenvsetTF}{mmm}
 {
  \septatrix_if_setenv:nTF { #1 } { #2 } { #3 }
 }

\ExplSyntaxOff

\newif\ifshowtodo

\newif\ifshowrem

\newif\ifshowrt

\newif\ifshowmq

\newif\ifverbosetoc

\newif\ifdebug
\debugtrue

\newif\iflocal
\localtrue

\ifenvsetTF{PAPER_DEBUG}{\debugtrue\localfalse}{}
\ifenvsetTF{PAPER_RELEASE}{\debugfalse\localfalse}{}

\ifdebug
	\showtodotrue
	\showremtrue
	\showrttrue
	\showmqtrue
	\verbosetoctrue
\else
	\iflocal
	\showtodotrue
	\showremfalse
	\showrtfalse
	\verbosetocfalse
	\showmqfalse
	\else
	\showtodofalse
	\showremfalse
	\showrtfalse
	\verbosetocfalse
	\showmqfalse
	\fi
\fi

\ifverbosetoc
\fi

\newcounter{rt}[section]

\newcounter{mq}[section]

\newcommand{\xmark}{\ding{55}}%

\newcommand{\scref}[1]{supplementary \cref{#1}}%
\definecolor{LightGray}{rgb}{0.92,0.92,0.92}

\newcommand{\mm}{\si{\milli\meter}~}
\DeclareSIUnit\mm{\milli\meter}
\newcommand{\km}{\si{\kilo\meter}~}
\DeclareSIUnit\km{\kilo\meter}
\DeclareSIUnit\micron{\micro\meter}
\DeclareSIUnit\pb{\peta\byte}
\DeclareSIUnit\tb{\tera\byte}
\DeclareSIUnit\gb{\giga\byte}
\DeclareSIUnit\mb{\mega\byte}
\DeclareSIUnit\bit{Bit}
\DeclareSIUnit\ghz{\giga\hertz}
\DeclareSIUnit\px{px}
\DeclareSIUnit\vx{vx}

\newcommand{\chindex}{CHI}

\newcommand{\brainfmt}[1]{\texttt{B}#1}
\newcommand{\brain}[1]{\brainfmt{\num[minimum-integer-digits=2]{#1}}}

\newcommand{\rom}[1]{\uppercase\expandafter{\romannumeral #1\relax}}

\newcommand{\area}[1]{#1\xspace}

\graphicspath{{images/}}

\begin{document}

\title{CytoNet: A Foundation Model for the Human Cerebral Cortex at Cellular Resolution}

\author[1,2]{Christian Schiffer}
\author[1,2]{Zeynep Boztoprak}
\author[1,2,3]{Jan-Oliver Kropp}
\author[1]{Julia Thönnißen}
\author[4]{Katia Berr}
\author[4,5]{Hannah Spitzer}
\author[6]{Mathis Bode}
\author[6]{Thomas Lippert}
\author[1,3]{Katrin Amunts}
\author[1,2,7]{Timo Dickscheid}

\affil[1]{Institute of Neuroscience and Medicine (INM-1), Research Centre Jülich, Jülich, Germany}
\affil[2]{Helmholtz AI, Research Centre Jülich, Jülich, Germany}
\affil[3]{Cécile \& Oscar Vogt Institute for Brain Research, University Hospital Düsseldorf, Düsseldorf, Germany}
\affil[4]{Institute of Computational Biology, Computational Health Center, Helmholtz Munich, Munich, Germany}
\affil[5]{Institute for Stroke and Dementia Research (ISD), LMU University Hospital, LMU Munich, Germany}
\affil[6]{Jülich Supercomputing Centre, Research Centre Jülich, Jülich, Germany}
\affil[7]{Computer Vision, Institute for Computational Visualistics, University of Koblenz, Koblenz, Germany}

\date{}
\maketitle

\begin{abstract}
	Studying the cellular architecture of the human cerebral cortex is essential for understanding how the brain is organized from the micro to the macro level, and how it functions.
	However, investigating complex texture patterns in histological images using automatic methods that can be scaled across whole brains remains a challenge.
	Here we introduce CytoNet, a foundation model trained on 1 million unlabeled microscopic image patches from over 4,000 histological sections from nine postmortem brains, and evaluated on over 2,000 sections from five additional brains excluded from self-supervised pretraining.
	Using co-localization in the cortical sheet for self-supervision, CytoNet learns to encode complex cellular patterns into expressive and anatomically meaningful feature representations.
	CytoNet supports multiple downstream applications, including area classification, laminar segmentation, quantification of microarchitectural variation, and exploratory mapping of cortical subdivisions.
	Functional parcellation analyses provided parcellation-dependent evidence for links between cytoarchitecture and macroscale functional organization.
	Together, these results establish CytoNet as a unified framework for scalable analysis of cortical microarchitecture and for testing links between cellular architecture and structure--function organization in the human cerebral cortex.
\end{abstract}

\section{Introduction}\label{sec:introduction} 

The cerebral cortex of the human brain is divided into areas with different functions, which exhibit complex hierarchical relationships and are structurally and functionally interconnected.
This forms the basis for a wide range of functions underlying cognition and consciousness.
Modern neuroscience has addressed the organization of the cerebral cortex in the living human brain with multimodal imaging~\citep{Glasser2016,Mathis2024}.
Three-dimensional reconstruction from serial sections is key to capture complementary organizational principles across scales.
For example, the \emph{BigBrain} project~\citep{Amunts2013} has processed 7404 full-brain histological sections to generate a volumetric 3D reconstruction at \SI{20}{\micron} resolution, approaching the cellular level.
Cytoarchitecture is defined by a regionally specific spatial organization of cells into distinct layers, columns, and areas, providing a mechanistic bridge linking multimodal measurements to anatomically well-defined structural entities~\citep{Campbell1904,Hilgetag2019,Amunts2022a}.
This cytoarchitectonic framework has guided cortical mapping for more than a century, from early area maps~\citep{Brodmann1909,VonEconomo1925,Vogt1919} to modern brain atlases~\citep{Zilles2010,Amunts2015,Ding2016,Amunts2020,Casamitjana2025}.
Advances in high-throughput microscopy have enabled whole-brain histological imaging at micrometer resolution, producing terabyte-scale datasets comprising thousands of sections from multiple human brains~\citep{Amunts2020}.
While acquiring such massive and detailed datasets is essential for decoding brain structure and function~\citep{Amunts2024}, their reconstruction and analysis demands scalable computational methods~\citep{Amunts2021}.

Advances in artificial intelligence (AI) suggest a route to address this challenge through large-scale foundation models that learn reusable representations from data~\citep{Bommasani2022}.
Such models have enabled breakthroughs ranging from protein structure prediction~\citep{Jumper2021} to vision~\citep{Radford2021,Oquab2024,Simeoni2025,Kirillov2023} and language~\citep{Chowdhery2023,OpenAI2024}.
A key driver of these successes is self-supervised learning, which extracts expressive features from massive unannotated datasets by generating implicit training signals from the data itself~\citep{He2020,Chen2020a}.
In this paradigm, computational approaches shift from designing specialized analysis pipelines to learning general feature representations that support data-driven investigation, prediction, and discovery.
In biomedical imaging, related pretrained models are emerging for brain MRI~\citep{Tak2026}, fMRI~\citep{Dong2024}, computational pathology~\citep{Chen2024,Xu2024,Zimmermann2024,Ding2025}, and cellular microscopy segmentation~\citep{Archit2025,Pachitariu2025}.
This motivates the development of foundation models for cellular-resolution whole-brain cytoarchitecture.

Related approaches in neuroscience also aim to reduce complex cortical organization to compact feature representations.
Gradient and embedding methods have shown that structural, functional, and microstructural similarity can reveal major axes of cortical organization~\citep{Margulies2016,Wagstyl2018,Paquola2019,Royer2022,Wang2025}.
These approaches typically start from predefined descriptors, such as laminar intensity profiles, receptor measurements, MRI-derived microstructure, or functional connectivity, and summarize their relationships into low-dimensional axes.
Foundation models address a complementary problem: they can learn a high-dimensional representation directly from image data, retaining local image features before any task-specific prediction or dimensionality reduction is applied.
This representation can serve directly as input to downstream analyses, including gradient-based approaches that identify dominant axes of variation and relate them to established anatomical, microstructural, or functional dimensions.

We introduce CytoNet, a foundation model trained by self-supervision on one million microscopic patches from over 4,000 cell-body-stained sections across nine postmortem brains and evaluated on over 2,000 sections from five brains excluded from pretraining.
CytoNet embeds local histological patterns in a shared feature space for cross-brain evaluation and adaptation.
Its SpatialNCE objective uses anatomical proximity within the cortical sheet to assign continuous distance-weighted similarity between unlabeled patches, avoiding manual annotation.
The learned representations capture shared cytoarchitectonic organization and brain-specific variation across brains and spatial scales.
They allow classifying over 100 Julich Brain Atlas areas~\citep{Amunts2020}, outperforming training from scratch, general self-supervision~\citep{Chen2020a}, supervised contrastive learning~\citep{Schiffer2021}, and vision or pathology foundation models~\citep{Oquab2024,Chen2024,Xu2024,Zimmermann2024}.
They also support label-efficient cortical layer segmentation, producing reliable laminar maps from few annotated samples.
Beyond mapping, CytoNet predicts structural variables, including layer-wise cell densities, better than established profile-based descriptors~\citep{Wagstyl2022}.
To test whether microarchitecture reflects systems-level specialization~\citep{Mesulam1998,Hilgetag2016}, we decode functional network parcellations under spatially aware validation and provide a proof-of-concept for exploratory mapping of cortical subdivisions.
CytoNet provides a scalable framework for learning, quantifying, and comparing cytoarchitecture across thousands of sections and whole brains.

\section{Results}\label{sec:results} 

We first describe CytoNet's self-supervised pretraining and examine how its feature space reflects local, global, and interindividual cytoarchitectonic variation.
We then evaluate structural prediction, functional decoding, cortical area and layer mapping, and exploratory parcellation.

\subsection{CytoNet learns cortical cytoarchitecture from spatial proximity}\label{sub:results_ai} 

\begin{figure}[t]
	\begin{center}
		\includegraphics[width=1.0\textwidth]{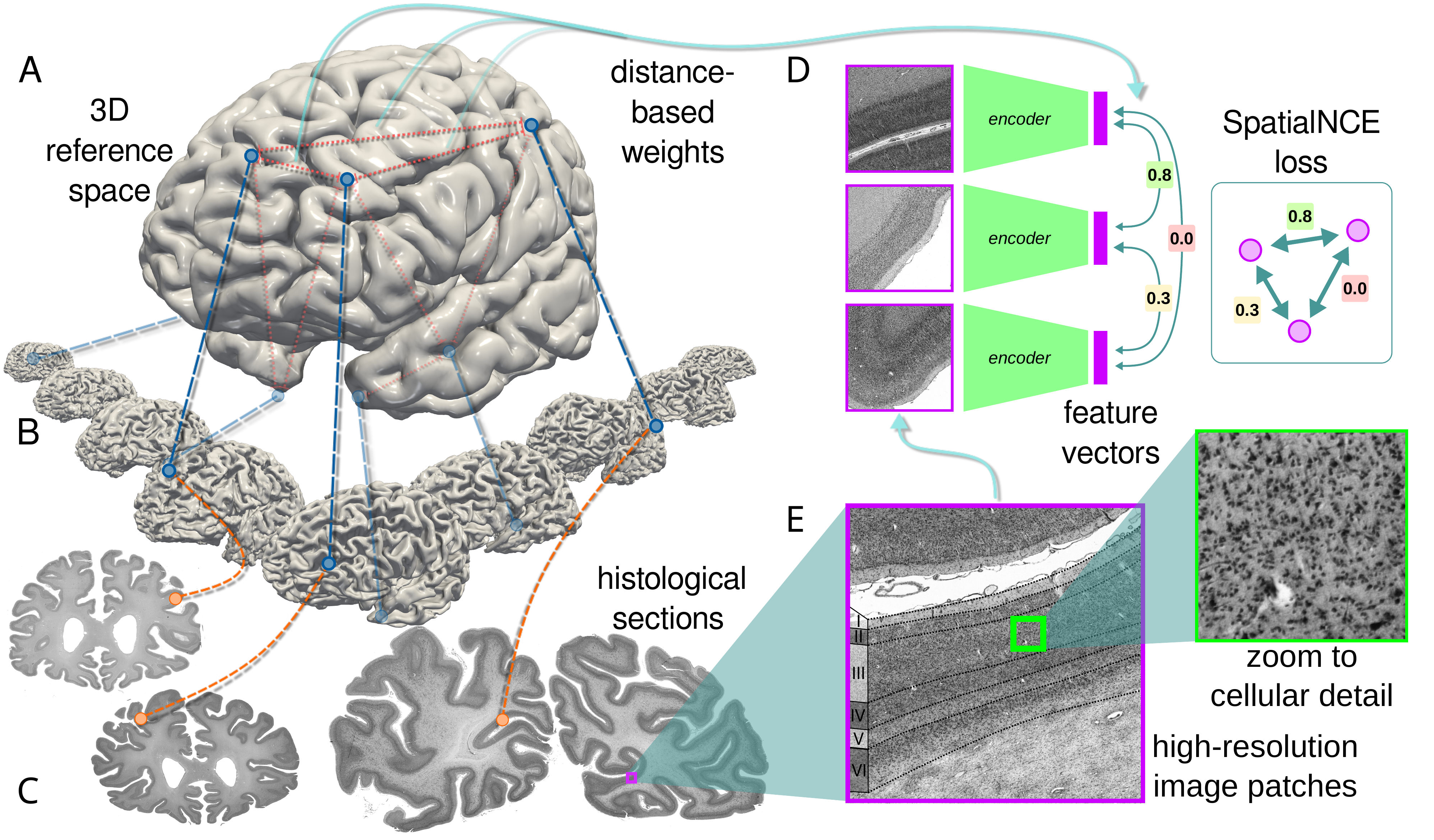}
	\end{center}
	\caption{
		\textbf{Self-supervised pretraining workflow using the SpatialNCE loss in CytoNet.}
		Spatial transformations between the MNI Colin 27~\citep{Holmes1998} 3D reference coordinate space (A) and microscopic scans of histological brain sections (C) of postmortem human brains (B) link high-resolution microscopic image patches (E) with corresponding approximate 3D locations in a common reference space.
		These locations allow distances between sampled image patches to be estimated within and across brains.
		SpatialNCE converts these distances into continuous similarity weights (D): nearby patches receive high weights and distant patches receive low weights.
		During training, CytoNet learns image features that reflect these spatially defined similarities, not knowing about the underlying 3D coordinates and using only image patches as model input.
	}
	\label{fig:schema}
\end{figure}

CytoNet encompasses a family of deep neural networks pretrained using self-supervised learning to map cortical image patches to high-dimensional feature vectors that capture cytoarchitectonic characteristics~(\cref{fig:schema}).
Pretraining is the initial stage in which the network learns a general image representation before being adapted to a specific downstream task.
The pretrained network ---referred to as \emph{encoder}--- can then be finetuned for downstream tasks such as area classification, layer segmentation, or structural prediction.
We use the name \emph{CytoNet} to refer to both the pretrained encoders and specialized finetuned downstream models.
Contrastive learning trains an encoder by defining which samples should have similar feature vectors and which samples should be separated in feature space.
The central question for cortical histology is therefore how to define similarity between unlabeled image patches.

In natural images, similar pairs for contrastive learning are often created by augmentations (e.g., random image transformations) that preserve semantic similarity, while disrupting irrelevant features.
For the type of data used for CytoNet, this assumption is problematic: common transformations can alter cytoarchitectonic structure, while confounding features such as blood vessels or folding geometry often remain unchanged~\citep{Kugelgen2021,Tian2020}.
Supervised contrastive learning~\citep{Khosla2020,Schiffer2021} circumvents this issue by using labels, but is limited by annotation cost.

We propose SpatialNCE, a continuous distance-weighted contrastive objective that encourages the model to map patches from nearby cortical locations to similar representations.
This leverages the anatomical continuity of the cortex as a heuristic for defining similarity, enabling scalable training on large unlabeled datasets without augmentations or manual annotations.
SpatialNCE builds on InfoNCE~\citep{Oord2019}, the basis of many modern contrastive methods~\citep{Chen2020a,He2020,Zbontar2021,Caron2020,Bardes2022}, and uses distances in a shared coordinate space as a proxy for semantic similarity.
Related spatial contrastive approaches use proximity to select discrete positive pairs, neighborhoods, triplets, or spatially guided targets~\citep{Jean2018,Oberstrass2024c,Le2024,Lin2023,Liu2021a,Robles2026}.
In comparison, SpatialNCE uses the full pairwise distances within a batch to assign continuous weights to all non-identical sample pairs.

For this purpose, square patches (\SI{2048}{\px} at \SI{2}{\micron\per\px}, approximately \SI{4}{\mm} per side) were sampled along the cortex from the histological brain sections~(\cref{fig:schema}).
All sections were co-registered to the MNI Colin 27 single subject reference space~\citep{Holmes1998}, enabling consistent distance computation across brains.
Each patch was assigned to a 3D coordinate in the MNI Colin 27 reference space through an approximate histological reconstruction workflow, enabling distances between patches to be computed within and across brains.
Five brains (\brain{2}, \brain{9}, \brain{11}, \brain{13}, and \brain{14}) were excluded from self-supervised pretraining and reserved for feature-space analyses and holdout validations.
Given a batch of image patches $x_i$ with corresponding 3D coordinates $p_i$ and normalized neural network features $z_i=f(x_i)$, the loss for patch $i$ is:
\begin{equation}
	\mathcal{L}_{i} = - \frac{1}{\sum_{j \neq i} \omega_{ij}} \sum_{j \neq i} \omega_{ij} \log \frac{ \exp\left( z_i^\top z_j / \tau \right) }{\sum_{k \neq i} \exp\left( z_i^\top z_k / \tau \right)},
\end{equation}
with temperature $\tau$ and similarity weights defined by the radial basis function (RBF) kernel
\begin{equation}
	\omega_{ij} = \exp\left(- \frac{||p_i - p_j||^2}{2\sigma^2} \right)
\end{equation}
with bandwidth $\sigma$.
The bandwidth $\sigma$ controls how the similarity weight decreases with anatomical distance.
Importantly, spatial coordinates define the training signal, but are never provided as model input.
CytoNet must therefore infer spatially consistent cytoarchitectonic features from image content alone.

SpatialNCE can be combined with any neural network architecture.
We evaluated ResNet50~\citep{He2016} and hybrid ResNet50-ViT-B~\citep{Dosovitskiy2020} models with modified input layers for large image patches.
Training processed up to \SI{4}{\tb} per epoch and \SI{600}{\tb} per run using either 16 compute nodes with 64 NVIDIA A100 GPUs or eight compute nodes with 32 NVIDIA GH200 superchips.
Runtime was up to 28 hours.
The downstream scaling analysis showed that model capacity and dataset size acted jointly, with larger pretraining datasets providing the clearest gains for higher-capacity architectures~(\cref{fig:results_area_classification}).

\subsection{CytoNet encodes cytoarchitectonic organization}\label{sub:results_features} 

\begin{figure}[tp]
	\begin{center}
		\includegraphics[width=0.96\textwidth]{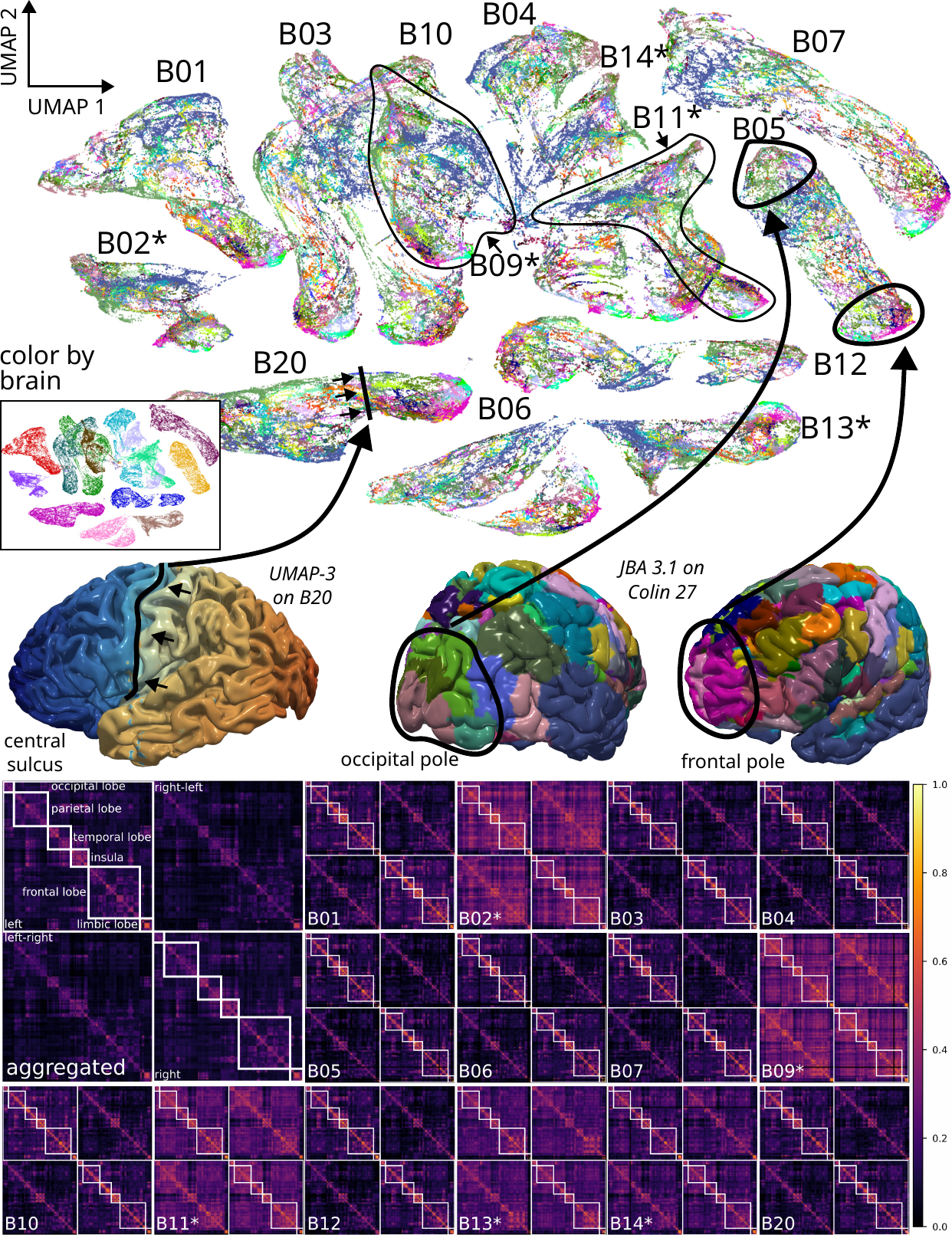}
	\end{center}
	\caption{
		\textbf{Anatomical plausibility of feature representations learned by CytoNet-R50-ViT (1M).}
		\textbf{Top:} 2D UMAP plot of the learned latent space, color coded by maximum probability labels of corresponding coordinates in the Julich Brain atlas (version 3.1,~\citet{Amunts2020}), as an approximate assignment to brain areas.
		A gap along the anterior-posterior axis aligns with the central sulcus, marking a prominent structural and functional division.
		Clusters corresponding to brains excluded from self-supervised pretraining (\brain{2}, \brain{9}, \brain{11}, \brain{13}, \brain{14}, marked by$^*$) appear more compact than the pretraining-brain clusters, but show a comparable cytoarchitectonic organization.
		\textbf{Bottom:} Aggregated pairwise cosine similarity of features across evaluated brains.
		Cosine similarity was computed between feature vectors from image patches, grouped by Julich Brain Atlas labels and averaged over all area pairs.
		Rows and columns represent brain areas, ordered by hemisphere, lobe, and label; area names are omitted for clarity (see \scref{tab:julich_brain_areas}).
	}\label{fig:umap_overview}
\end{figure}

We examined how CytoNet-R50-ViT (1M) organizes cytoarchitectonic information in its feature space (\cref{fig:umap_overview}, top\footnote{Interactive versions of selected figures are available at \url{https://go.fzj.de/cytonet-interactive}.}).
CytoNet-R50-ViT (1M) is a hybrid ResNet50–ViT-B model pretrained on one million cortical patches (see \cref{sub:self_supervised_pretraining} for details).
2D UMAP embeddings~\citep{McInnes2018} of patches from all 14 brains revealed distinct brain-specific manifolds (Calinski--Harabasz index, \chindex{} = $3352.93$; \citealp{Calinski1974}) with consistent internal organization: atlas labels indicating different cytoarchitectonic areas~\citep{Amunts2020} clustered coherently (\chindex{} = $738.31$).
The UMAP embeddings consistently showed a subdivision of clusters at the central sulcus, which is an important anatomical landmark separating motor and somatosensory areas.
Brains excluded from self-supervised pretraining (\brain{2}, \brain{9}, \brain{11}, \brain{13}, and \brain{14}) formed more compact clusters than the pretraining brains.
Their mean within-brain feature spread, measured as the root-mean-square standard deviation across feature dimensions, was $0.307\pm 0.018$, compared with $0.396\pm 0.008$ for the pretraining brains ($-22.5\%$).
Despite their reduced spread, these brains retained comparable internal structure.
An extended embedding analysis is provided in \scref{sub:additional_embedding_analyses}.

To assess cross-brain cytoarchitectonic similarity, we mean-aggregated pairwise feature cosine similarities by atlas area~(\cref{fig:umap_overview}, bottom).
The resulting matrices showed higher similarity within than between areas, and their patterns were highly correlated across brains (Pearson $r = 0.82\pm 0.09$), indicating stable inter-area relationships.
Brains excluded from self-supervised pretraining showed higher overall similarity and weaker block structure, consistent with their reduced specificity in the UMAP embeddings.
Jointly fitted CytoNet PCA components showed reproducible surface gradients in brains included in and excluded from pretraining, providing an anatomical view of feature-space variance~(\scref{fig:pca_on_mesh_comparison}).

\begin{figure}
	\begin{center}
		\includegraphics[width=1.0\textwidth]{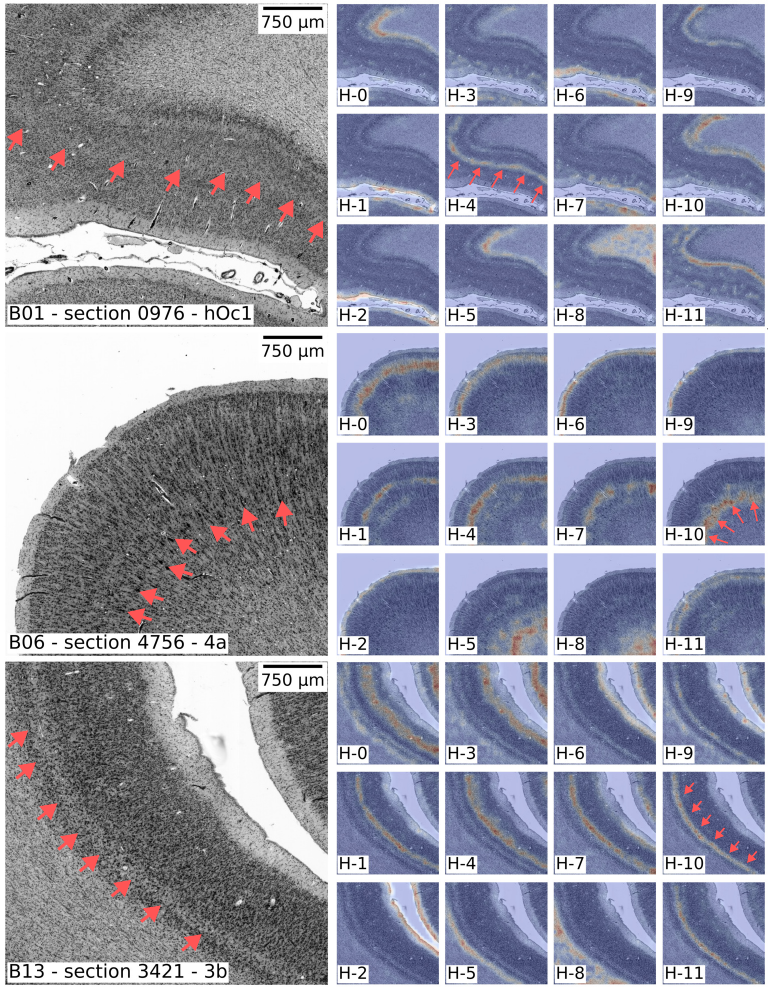}
	\end{center}
	\caption{
		\textbf{Local ViT attention maps for CytoNet-R50-ViT (1M).}
		Example patches are shown from \area{hOc1} in \brain{1}, \area{4a} in \brain{6}, and \area{3b} in \brain{13}.
		\brain{13} was excluded from self-supervised pretraining.
		The areas correspond to primary visual cortex~\citep{Amunts2000}, primary motor cortex~\citep{Geyer1996}, and primary somatosensory cortex~\citep{Geyer1999}.
		Each row shows the input image on the left and class-token attention scores from all 12 heads of the first transformer layer overlaid on the image on the right, with red indicating stronger attention.
		Arrows highlight cytoarchitectonic landmarks: the stripe of Gennari in primary visual cortex, Betz giant cells in layer V of motor cortex, and a pronounced layer IV in somatosensory cortex.
		Attention scores were gamma transformed ($\gamma = 0.5$) for visualization.
	}
	\label{fig:attention_overview}
\end{figure}

\begin{figure}
	\begin{center}
		\includegraphics[width=0.95\textwidth]{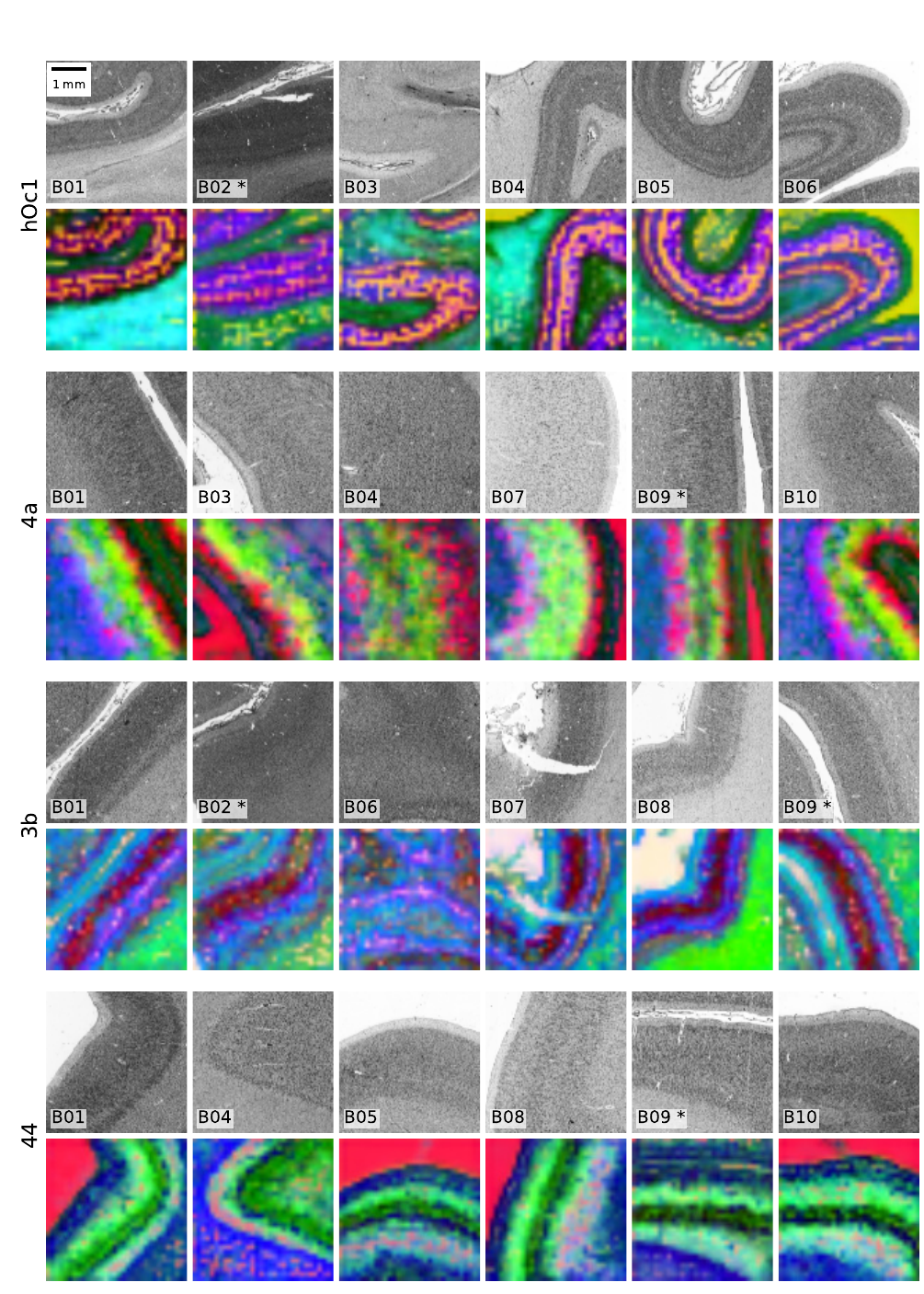}
	\end{center}
	\caption{
		\textbf{DINO-style patch-token PCA/RGB visualization for CytoNet-R50-ViT (1M).}
		For each displayed area, the class token was removed and a three-component PCA was fitted to spatial patch-token embeddings from all displayed samples of that area.
		The first three PCA components were percentile-normalized within each area and encoded as RGB channels, following DINO-style token visualizations~\citep{Caron2021}.
		Colors are comparable within each area block and should not be compared across different areas.
		Brains marked with ${}^*$ were excluded from self-supervised pretraining.
	}\label{fig:token_pca_by_area}
\end{figure}

First-layer class-token attention maps for CytoNet-R50-ViT (1M) highlighted recognizable cytoarchitectonic landmarks, including the stripe of Gennari, Betz giant cells, and the pronounced layer IV of primary somatosensory cortex~(\cref{fig:attention_overview}, see also \cref{sub:local_vit_visualization}).
Patch-token PCA/RGB visualizations across areas and brains provided complementary views of spatial feature organization~(\scref{fig:token_pca_by_area}).
These visualizations show that CytoNet attends to laminar patterns, tissue boundaries, and characteristic cellular structures.
The composition and spatial arrangement of these features define cytoarchitectonic areas, making them directly useful for cortical layer segmentation and area classification.


\subsection{Applications of CytoNet}\label{sub:applications}

To demonstrate the utility of CytoNet feature representation for human brain mapping, we evaluated structural prediction, functional decoding, area and layer mapping, and exploratory parcellation.

\subsubsection*{Predicting structural variation in cytoarchitecture}

\begin{figure}
	\begin{center}
		\includegraphics[width=1.0\textwidth]{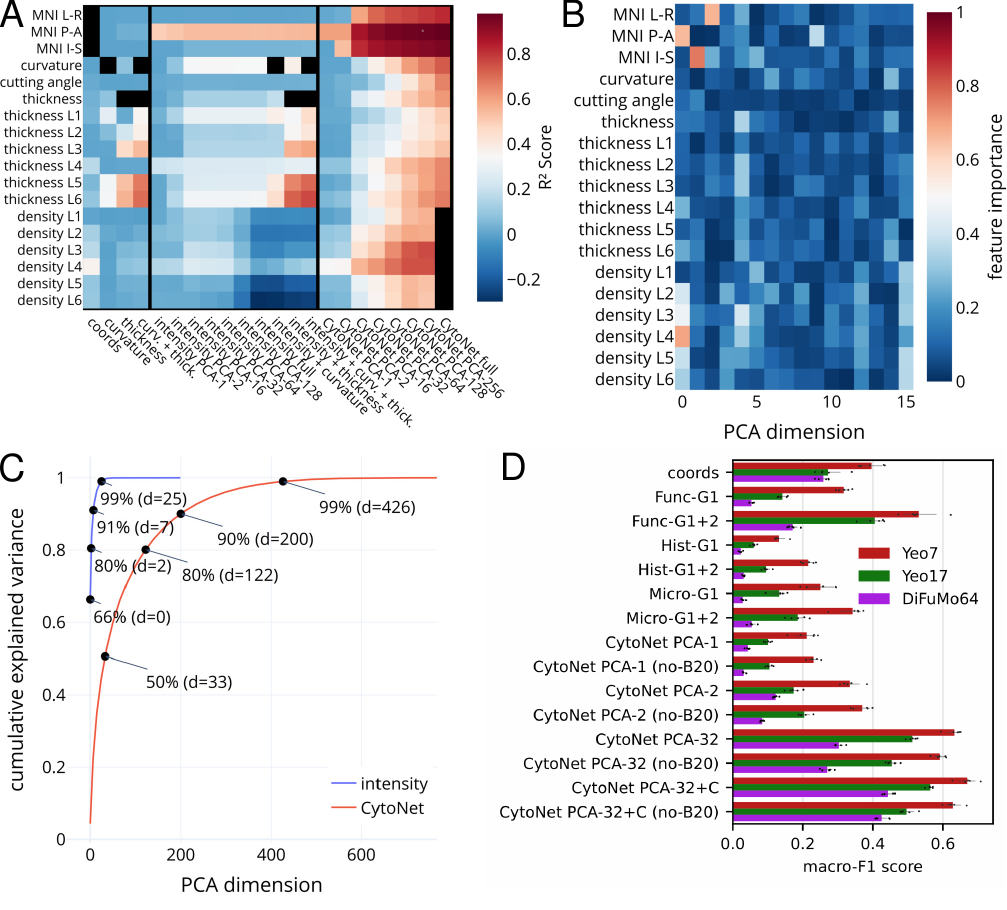}
	\end{center}
	\caption{
		\textbf{Decoding structural properties and functional parcellations from CytoNet representations in \brain{20}.}
		\textbf{A:} Linear regression models using varying subsets of PCA components compared CytoNet features with intensity profiles, curvature, cortical thickness, BigBrain-space coordinates, and intensity profiles concatenated with cortical curvature, cortical thickness, or both.
		Reported values reflect the $R^2$ across 5-fold cross-validation.
		CytoNet PCA-$N$ indicates the first $N$ PCA components of CytoNet features.
		Black cells indicate invalid combinations, either because the input feature set contains the prediction target itself or because the feature dimensionality precluded fitting the regression model.
		\textbf{B:} Absolute feature importance scores derived from regression coefficients for the first 16 PCA components of CytoNet features showed that components 1–3 were strongly associated with spatial location in MNI space and the density of cortical layer IV, while other properties were predominantly encoded in higher components.
		I to VI denote cortical layers I to VI.
		\textbf{C:} Cumulative explained variance across PCA components indicates that CytoNet features capture substantially more variance than intensity profiles.
		\textbf{D:} Decoding functional parcellation labels (Yeo7, Yeo17; \citet{ThomasYeo2011}, and DiFuMo64; \citet{Dadi2020}) from different feature sets using multinomial logistic regression.
		Reported values reflect macro-F1 across 5-fold cross-validation.
		PCA-$N$ denotes the first $N$ CytoNet principal components, so PCA-1 and PCA-2 provide the low-dimensional comparison to G1 and G1+G2 gradient baselines.
		CytoNet PCA-32 outperformed histological (Hist-G, \citet{Paquola2019}) and microstructural (Micro-G, \citet{Royer2022}) gradient baselines as well as a low-dimensional functional-gradient reference (Func-G, \citet{Margulies2016}).
		A baseline using only fs\_LR vertex coordinates quantifies the contribution of cortical location.
		CytoNet-R50-ViT (1M, no-B20) was not pretrained on \brain{20}.
		Concatenating CytoNet features with coordinates (CytoNet PCA-32+C) yields the highest score, indicating complementary information beyond spatial position.
	}\label{fig:results_structure_prediction}
\end{figure}

We assessed morphological and cytoarchitectonic properties encoded by CytoNet-R50-ViT (1M) and CytoNet-R50-ViT (1M, no-B20) using variables extracted from the BigBrain dataset \brain{20}~\citep{Amunts2013}.
The first model included \brain{20} during self-supervised pretraining, whereas the second did not.
We compared CytoNet features with BigBrain intensity profiles~\citep{Wagstyl2018}, which are established descriptors of cortical microstructure and macroscale organization~\citep{Wei2018,Wagstyl2018,Paquola2019}, both alone and combined with cortical curvature, cortical thickness, or both.

Both intensity profiles and CytoNet features predicted anterior-posterior position well, which may partly reflect section-wise staining or processing gradients because this axis follows the histological sectioning direction in BigBrain.
CytoNet additionally captured all three spatial axes, cortical thickness, curvature, and cutting angle~(\cref{fig:results_structure_prediction}, A).
Adding curvature and thickness to intensity profiles improved prediction of layer thicknesses ($R^2=0.28$--$0.76$ versus $0.13$--$0.32$), while CytoNet achieved comparable or higher accuracy for several layers ($R^2=0.48$--$0.67$).
For layer-wise cell densities, the geometry-augmented intensity profiles remained weak ($R^2=-0.25$--$0.12$), whereas CytoNet PCA--32 reached $R^2=0.28$--$0.62$.
A coordinate-only baseline explained little variance in non-spatial targets, except for modest prediction of layer IV density ($R^2=0.36$) and layer IV thickness ($R^2=0.16$).
Excluding \brain{20} from pretraining largely preserved structural-property prediction but strongly reduced explicit left-right coordinate encoding, for example from $R^2=0.71$ to $0.02$ for PCA--32.
Higher-dimensional CytoNet projections generally improved predictive accuracy, whereas intensity-profile projections remained largely constant~(\cref{fig:results_structure_prediction}, A).
Consistent with this, cumulative explained variance was markedly higher for CytoNet features than for intensity profiles~(\cref{fig:results_structure_prediction}, C).

\subsubsection*{Decoding functional organization from cytoarchitecture}

We trained multinomial logistic regression models to decode Yeo7 and Yeo17 networks~\citep{ThomasYeo2011} and DiFuMo64 modes~\citep{Dadi2020} from CytoNet representations (\cref{fig:results_structure_prediction}, D).
At 32 PCA components, CytoNet strongly predicted functional network identity and exceeded the fs\_LR coordinate baseline~\citep{VanEssen2012}, increasing macro-F1 from $0.40 \pm 0.03$ to $0.63 \pm 0.02$ for Yeo7, from $0.27 \pm 0.04$ to $0.51 \pm 0.01$ for Yeo17, and from $0.26 \pm 0.02$ to $0.30 \pm 0.01$ for DiFuMo64.
Performance increased systematically with feature dimensionality.
PCA--1 achieved macro-F1 scores of $0.21 \pm 0.04$, $0.10 \pm 0.01$, and $0.04 \pm 0.01$ for Yeo7, Yeo17, and DiFuMo64, respectively, while PCA--2 achieved $0.33 \pm 0.03$, $0.17 \pm 0.02$, and $0.12 \pm 0.01$.
PCA--32 achieved the values reported above.
CytoNet-R50-ViT (1M, no-B20) retained most of this performance at PCA--32, reaching $0.59 \pm 0.02$, $0.45 \pm 0.02$, and $0.27 \pm 0.02$, respectively, indicating that decoding did not strongly depend on exposure to \brain{20} during pretraining.

We next tested whether CytoNet's improvement over coordinates could be explained by the inherent spatial smoothness of microstructural and functional organization alone.
Using a spin-based spatial null on the parcellation labels in fs\_LR space, the CytoNet--coordinate advantage was strongest for Yeo17 ($\Delta=0.250$, 95\% CI $[0.231, 0.274]$, one-sided $p=0.0028$, Holm--Bonferroni-corrected $p=0.0084$).
For Yeo7, the observed improvement was similarly large but did not remain significant after correction ($\Delta=0.230$, 95\% CI $[0.216, 0.242]$, $p=0.0357$, corrected $p=0.0714$).
For DiFuMo64, the smaller improvement over coordinates was not supported by the spatial null ($\Delta=0.059$, 95\% CI $[0.037, 0.078]$, $p=0.951$, corrected $p=0.951$).
CytoNet-R50-ViT (1M, no-B20) showed the same parcellation-dependent pattern.
Yeo17 remained significant after correction ($\Delta=0.182$, 95\% CI $[0.157, 0.209]$, $p=0.0038$, corrected $p=0.0114$), whereas Yeo7 ($\Delta=0.194$, 95\% CI $[0.165, 0.227]$, $p=0.0424$, corrected $p=0.0848$) and DiFuMo64 ($\Delta=0.020$, 95\% CI $[-0.005, 0.044]$, $p=0.982$, corrected $p=0.982$) did not.
Thus, although absolute decoding performance was highest for the coarser Yeo7 parcellation, the strongest evidence for cytoarchitectonic information beyond spatial smoothness was obtained for the finer-grained Yeo17 organization.

CytoNet PCA--32 outperformed histological gradients~\citep{Paquola2019}, microstructural gradients~\citep{Royer2022}, and the first two functional-gradient axes~\citep{Margulies2016} across all three parcellations.
This indicates that CytoNet captures combinations of cytoarchitectonic features that align with functional networks but are not represented by a few dominant gradient axes.
Cortical position provided complementary information, increasing macro-F1 from $0.63$ to $0.67$ for Yeo7, from $0.51$ to $0.56$ for Yeo17, and from $0.30$ to $0.44$ for DiFuMo64 when added to CytoNet PCA--32.
The same pattern occurred for CytoNet-R50-ViT (1M, no-B20), whose combined representation reached $0.63 \pm 0.03$, $0.50 \pm 0.02$, and $0.43 \pm 0.02$, respectively.

\subsubsection*{Mapping of brain areas and cortical layers}

We next evaluated whether CytoNet features support explicit mapping of cytoarchitectonic organization in supervised tasks.
We distinguish three settings: Julich Brain Atlas~\citep{Amunts2020} area classification within the atlas framework used for model development, Allen Adult Human Brain Atlas~\citep{Ding2016} area classification as an external transfer experiment, and cortical layer segmentation within histological image patches.

\begin{figure}
	\begin{center}
		\includegraphics[width=1.0\textwidth]{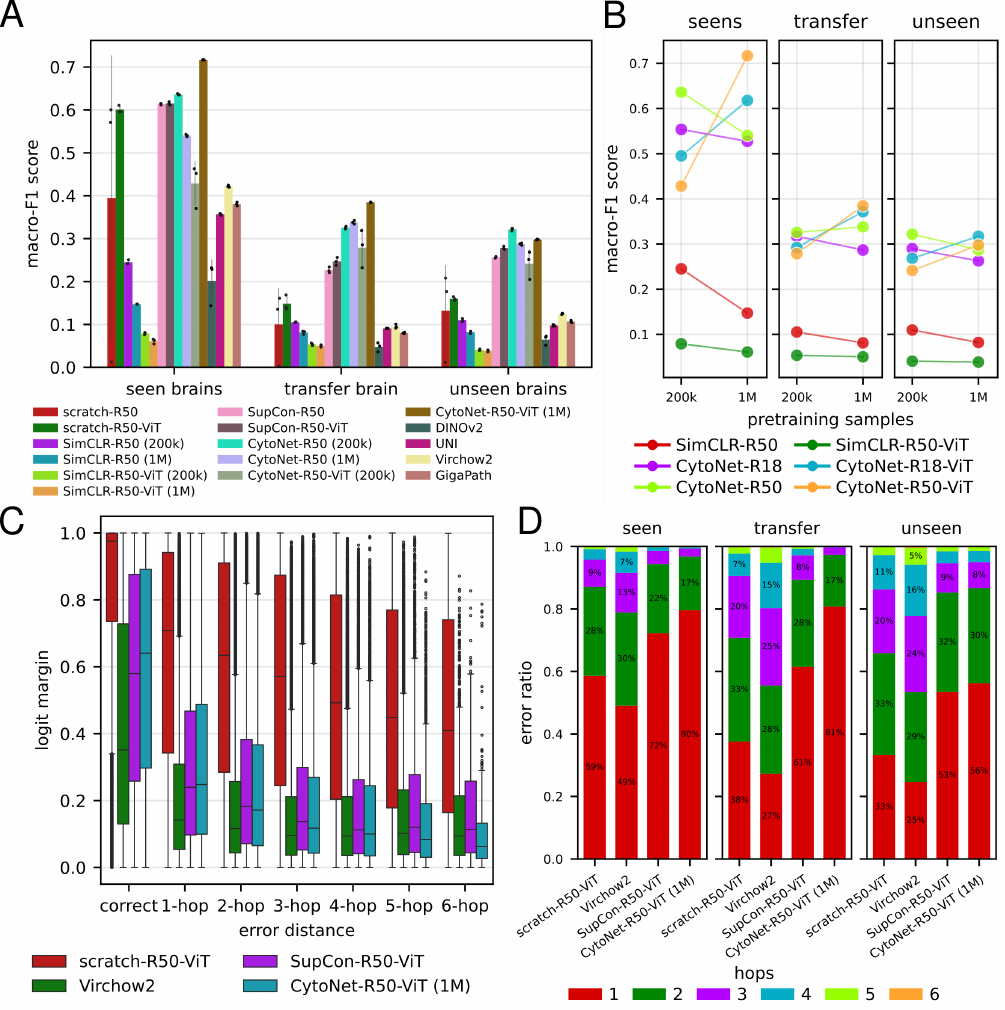}
	\end{center}
	\caption{
		\textbf{Performance for classifying 113 cytoarchitectonic areas from the Julich Brain Atlas~\citep{Amunts2020}.}
		\textbf{A:}~Macro-F1 scores obtained by linear probing of different models across seen, transfer, and unseen brains.
		The number of pretraining samples is indicated behind the model name, if applicable.
		Black dots show results of three repeated runs; black bars show standard deviations across runs.
		See \scref{tab:scores_table,tab:cytonet_variants_table} for more detailed scores and variant controls.
		\textbf{B:} Scaling analysis of SimCLR and CytoNet models with varying model sizes and numbers of self-supervised pretraining samples.
		CytoNet benefits from larger models when more pretraining samples are provided.
		\textbf{C:}~Logit margins for predictions stratified by prediction correctness and error distance for CytoNet-R50-ViT (1M), scratch-R50-ViT, SupCon-R50-ViT, and Virchow2.
		SimCLR is omitted due to overall low performance.
		Logit margins ---the difference between the top two logits--- serve as a proxy for model confidence and distance to the decision boundary~\citep{Ngnawe2024}.
		Correct predictions reveal higher confidence, while incorrect predictions show decreasing confidence with increasing error distance.
		\textbf{D:}~Distribution of prediction errors by error distance for the same models across seen, transfer, and unseen brains.
		Error distance was defined as the number of hops between the predicted and true brain area in the adjacency graph of the Julich Brain Atlas 3.1~\citep{Amunts2020}, where 1-hop errors correspond to directly adjacent areas, and larger distances reflect increasing topological separation.
	}\label{fig:results_area_classification}
\end{figure}

For Julich Brain Atlas area classification, CytoNet performance depended on architecture and pretraining scale, with the strongest variants outperforming models trained from scratch, SimCLR~\citep{Chen2020a}, supervised contrastive baselines~\citep{Schiffer2021}, and pretrained foundation-model baselines (\cref{fig:results_area_classification}, A and \scref{tab:scores_table}, see \cref{sub:brain_area_classification} for model details).
CytoNet-R50-ViT (1M), the strongest variant on seen and transfer brains, reached macro-F1 scores of $0.72\pm0.00$ on seen brains, $0.38\pm0.00$ on the transfer brain, and $0.30\pm0.00$ on unseen brains.
On unseen brains, CytoNet-R18-ViT (1M) and CytoNet-R50 (200k) reached the highest CytoNet score of $0.32\pm0.00$, compared with $0.28\pm0.00$ for the strongest supervised contrastive baseline and $0.12\pm0.00$ for the strongest pretrained foundation-model baseline.
The unseen-brain split contained \brain{2}, \brain{9}, \brain{11}, \brain{13}, and \brain{14}, enabling evaluation across brains excluded from both self-supervised pretraining and supervised classifier training.
Label-efficiency analyses on seen and unseen brains showed that adding even a small number of supervised samples per label substantially improved performance.
CytoNet performed best in this low-label regime, indicating that the pretrained representation can be adapted efficiently when limited annotations are available.

To assess target-brain adaptation, we continued self-supervised pretraining on \brain{9} for 10 epochs before training the area classifier.
Without \brain{9} labels, this increased macro-F1 on \brain{9} from $0.25$ to $0.47$ relative to CytoNet-R50-ViT (1M), which excluded \brain{9} from self-supervised pretraining.
With all available \brain{9} labels, performance increased from $0.69$ to $0.72$.
A geodesic-distance variant of CytoNet-R50 (200k), using cortical-surface distances in the SpatialNCE objective, performed below the corresponding Euclidean-distance model, with macro-F1 differences of $-0.050$, $-0.024$, and $-0.015$ on seen, transfer, and unseen brains, respectively~(\scref{tab:cytonet_variants_table}).
SimCLR variants performed poorly across all splits, often below training from scratch, and retrieval analysis showed that their features clustered images primarily by tissue morphology and vascular patterns rather than area identity~(\scref{fig:simclr_comparison}).

\begin{figure}
	\begin{center}
		\includegraphics[width=1.0\textwidth]{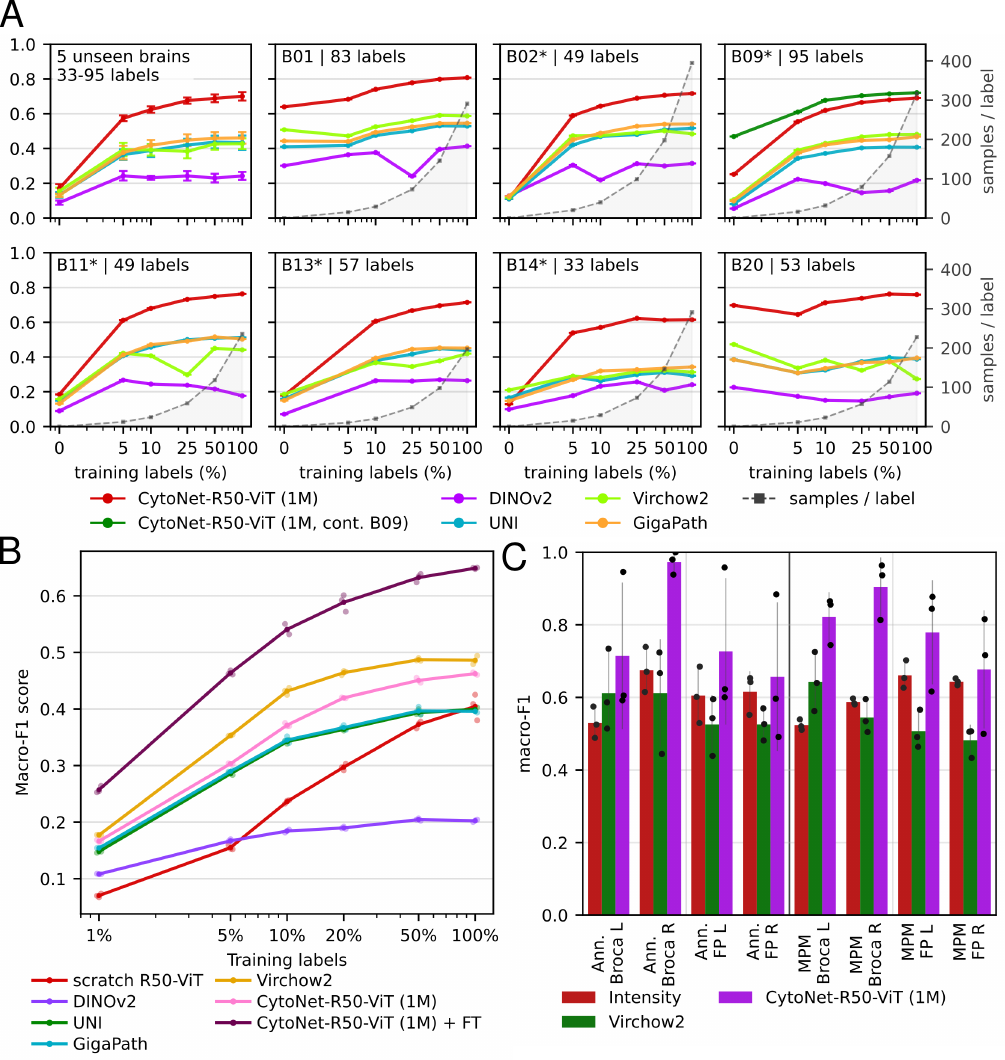}
	\end{center}
	\caption{
		\textbf{Label efficiency and clustering performance.}
		\textbf{A:}~Label efficiency analysis for cytoarchitectonic area classification in the Julich Brain Atlas.
		Pretrained models were linearly probed using a fraction of annotations in a specific brain.
		${}^*$ marks brains excluded from CytoNet pretraining.
		Panels indicate the number of labels available for linear probing in the respective brain.
		The first panel reports average performance across the five unseen brains.
		0\% denotes the default linear-probing setting: regular supervised training for seen brains and no target-brain labels for unseen brains; non-zero budgets use only the indicated fraction of labels from the respective target brain.
		Label sets differ by brain, so 0\% unseen-brain values in this panel are evaluated only over labels present in the corresponding target brain.
		\textbf{B:} Label efficiency for classification of neocortex areas in the Allen Adult Human Brain Atlas.
		Transparent markers show results of 5 individual runs.
		CytoNet-R50-ViT (1M) - FT was finetuned, all other models were linearly probed.
		\textbf{C:}~Clustering performance at the anatomical number of subdivisions for frontal pole (FP,~\citet{Bludau2014}) and Broca's regions (Broca,~\citet{Amunts1999}).
		Clustering was performed with $k=2$ K-means clusters; separately per brain, hemisphere, area, and pre-localization strategy; using intensity profiles~\citep{Wagstyl2018}, Virchow2~\citep{Zimmermann2024}, and CytoNet-R50-ViT (1M) features.
		Error bars show the standard deviation across brains; black dots show the mean over five randomly initialized K-means runs per brain.
		Example visualizations are provided in \scref{fig:clustering_overview}.
	}\label{fig:results_label_efficiency_clustering}
\end{figure}

The following error-distance analysis focused on the Julich Brain Atlas setting, where area adjacency was defined by the atlas topology.
CytoNet misclassifications were strongly concentrated near the correct area: approximately 80\% of errors in seen and transfer brains occurred within one adjacency hop and over 95\% within two hops~(\cref{fig:results_area_classification}, D).
For unseen brains, these proportions were lower, but errors remained topologically plausible.
Control models showed the same pattern more weakly, with SupCon-R50-ViT errors also enriched near the correct area and scratch-R50-ViT and Virchow2 producing more long-range errors.
Logit-margin confidence was higher for correct predictions and decreased systematically with error distance across CytoNet and the control models~(\cref{fig:results_area_classification}, C).
These findings indicate that CytoNet errors are concentrated at difficult atlas borders, consistent with uncertainties faced by human experts.

We next evaluated Allen Adult Human Brain Atlas area classification as an external transfer experiment across dataset, laboratory, and annotation ontology.
Because the Allen atlas contains one donor, the section-wise split tests transfer to unseen sections within an independently acquired atlas, while the five brains excluded from self-supervised pretraining provide a complementary cross-brain evaluation.
Finetuning CytoNet-R50-ViT (1M) achieved the strongest performance across all annotation budgets, increasing from macro-F1 $0.26\pm0.01$ with 1\% of the training samples to $0.65\pm0.00$ with the full training set~(\cref{fig:results_label_efficiency_clustering}, B; \scref{tab:allen_label_efficiency_scores}).
As a frozen feature extractor, CytoNet-R50-ViT (1M) reached $0.46\pm0.00$ with the full training set, comparable to but below Virchow2 ($0.49\pm0.01$).

\begin{figure}
	\begin{center}
		\includegraphics[width=1.0\textwidth]{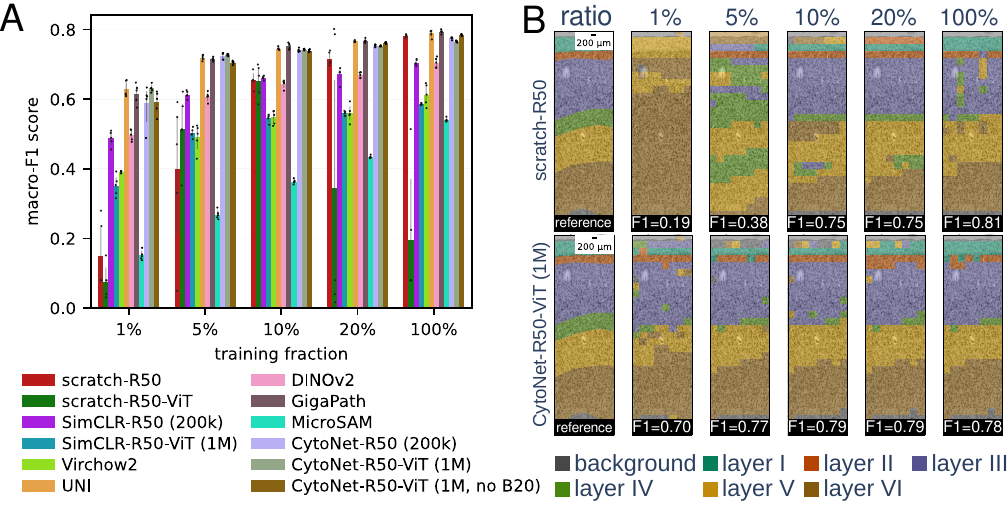}
	\end{center}
	\caption{
		\textbf{Results for cortical layer segmentation.}
		\textbf{A:} Macro-F1 scores from linear probing of different models with varying amounts of training data across five-fold cross-validation, evaluated on a dedicated test set (also see \scref{tab:layer_segmentation_scores}).
		If applicable, the number of pretraining samples is indicated after the model name.
		\textbf{B:} Example segmentation of cortical layers in area 44 (\citet{Amunts1999}) from the test set for scratch and CytoNet-R50-ViT (1M) models for increasing fractions of the training set.
	}\label{fig:results_layer_segmentation}
\end{figure}

Cortical layer segmentation further demonstrated the richness and data efficiency of CytoNet features.
Under linear probing, CytoNet-R50-ViT (1M) was competitive with the strongest pretrained foundation-model baselines in the low-label regime, reaching macro-F1 $0.63\pm0.01$ with 1\% of the training data and $0.73\pm0.00$ with 5\%~(\cref{fig:results_layer_segmentation}, A; \scref{tab:layer_segmentation_scores}).
Performance saturated around $0.77$--$0.78$, approaching UNI and GigaPath at full training data and exceeding SimCLR, Virchow2, DINOv2, and MicroSAM in this setup.
CytoNet-R50-ViT (1M, no-B20), which had not seen \brain{20} during self-supervised pretraining, showed similar linear-probing performance ($0.59\pm0.03$ at 1\%, $0.71\pm0.00$ at 5\%, and $0.78\pm0.00$ at 100\%), indicating that layer-segmentation performance transfers without prior self-supervised exposure to the same brain.
Finetuning improved some configurations at larger annotation budgets but was less stable than linear probing in small-data regimes and for transformer backbones, particularly for CytoNet-R50-ViT (1M)~(\scref{tab:layer_segmentation_scores}).

\subsubsection*{Data-driven exploration of brain areas}

We asked whether CytoNet features support exploratory refinement of brain parcellations through clustering.
Identifying and delineating brain areas remains a central challenge in brain mapping and analysis, particularly in regions lacking anatomical landmarks.
While historical cytoarchitectonic maps (e.g., \citet{Brodmann1909}) provide a foundational parcellation, many regions have since been identified based on refined histological evidence, often by subdividing existing areas.
This process is still ongoing, and even the most detailed contemporary atlases remain incomplete:
the Julich Brain atlas is published as a continuously growing collection of probabilistic maps, and parts of the cortex have not yet been charted~\citep{Amunts2020}.
Fully automatic discovery requires both a suitable representation and robust model selection for the number and geometry of clusters.

We therefore evaluated whether clustering at the known anatomical number of subdivisions recovers expert and atlas-defined subarea structure across two regions: frontal pole areas Fp1/Fp2~\citep{Bludau2014} and Broca's areas BA44/BA45~\citep{Amunts1999}.
The analysis included \brain{6}, \brain{9}, and \brain{20}, with clustering performed separately within each hemisphere for both expert-annotation and Julich Brain MPM pre-localizations.
We treated expert annotations as the primary validation against brain-specific cytoarchitectonic boundaries, while the MPM-based analysis served as an atlas-derived robustness check with known spatial uncertainty.
The resulting anatomical-$k$ macro-F1 scores are shown in \cref{fig:results_label_efficiency_clustering}C, with representative surface visualizations in \scref{fig:clustering_overview}.

CytoNet features outperformed both Virchow2 and BigBrain intensity-profile baselines in all eight region-by-hemisphere-by-pre-localization settings at the anatomical $k$.
Broca's region showed the strongest and most reproducible recovery, particularly in the right hemisphere.
We also tested whether the number of clusters could be selected directly from the data using silhouette score, Calinski--Harabasz index, Davies--Bouldin index, gap statistic, and bootstrap stability~(\scref{fig:clustering_k_selection_summary}).
These criteria did not converge on a single subdivision count across regions and brains, indicating that automatic cytoarchitectonic discovery still requires dedicated clustering and model-selection strategies, particularly for graded or hierarchical organization.
CytoNet therefore provides a substantially stronger representation for exploratory mapping, but does not by itself resolve unsupervised area discovery.



\section{Discussion}\label{sec:discussion} 

CytoNet learns cortical cytoarchitecture from large-scale histological data using self-supervision from anatomical proximity.
The model encodes laminar and areal organization into a reusable feature space that supports area classification, layer segmentation, structure prediction, exploratory parcellation, and functional decoding.
Across these applications, spatial continuity proves an effective training signal for capturing cytoarchitectonic organization at whole-brain scale.
CytoNet therefore provides a scalable pretrained representation for cortical histology and a methodological basis for studying cytoarchitectonic organization across many brains and sections.

The SpatialNCE training strategy assumes that spatially close samples are often semantically consistent.
In the cortex, nearby textures often represent architectural variations within the same cytoarchitectonic unit.
This distinguishes CytoNet from contrastive~\citep{Caron2020,Chen2020a,He2020,Khosla2020} and non-contrastive~\citep{Grill2020,Chen2021} self-supervised approaches based on image augmentations.
It also addresses the limitations observed with SimCLR~\citep{Chen2020a}, which sometimes underperformed scratch training by relying on augmentation-invariant cues such as vascular patterns or local curvature rather than cytoarchitecture.
This behavior is an example of shortcut learning~\citep{Geirhos2020}.
Augmentations implicitly determine which features constitute task-relevant \emph{content} and task-irrelevant \emph{style}~\citep{Kugelgen2021}.
In histology, standard intensity, noise, and geometric transformations can preserve confounds, whereas stronger transformations may distort cytoarchitectonic structure.
Large fields of view capture laminar profiles but include these confounds, while smaller inputs sacrifice important context.
Spatial proximity instead acts as a biologically grounded augmentation that uses cortical continuity across 3D space and subjects to preserve cytoarchitecture while disrupting confounds.
Related objectives have used proximity to define pairs, neighborhoods, or spatial targets in 3D polarized light imaging~\citep{Axer2011,Oberstrass2024c}, medical imaging~\citep{Le2024,Lin2023,Liu2021a,Robles2026}, and remote sensing~\citep{Jean2018}.
SpatialNCE differs by assigning continuous distance-dependent weights to every non-identical pair and integrating multiple brains through a common anatomical reference space.
The formulation may also apply to other spatially anchored domains.

Earlier cytoarchitectonic approaches incorporated spatial context differently.
\citet{Spitzer2018} aligned feature-space distances with geodesic distances along the cortical sheet of one reconstructed brain.
This direct alignment imprints cortical-sheet geometry onto the representation, whereas SpatialNCE uses anatomical distances as contrastive weights and can retain variation unexplained by location.
A graph-neural-network framework~\citep{Schiffer2021a} showed how learned image features can be combined with approximate 3D cortical topology for supervised classification.
CytoNet complements this framework because its pretrained features can be inserted into the same graph-based pipeline.
CytoNet builds on the aforementioned ideas by scaling spatially informed feature learning to multi-brain whole-cortex pretraining in a shared anatomical reference space.

CytoNet's training objective defines similarity from the Euclidean distance between patch locations, but spatial coordinates are never provided as model inputs.
The learned feature space nevertheless aligns strongly with anatomical location, indicating that local image information reflects the spatial continuity of cortical organization.
At the same time, the objective remains flexible enough to deviate from this imposed geometry, for example by separating motor and somatosensory areas across the central sulcus despite their adjacency in 3D space.
Notably, cortical-surface distances performed worse than approximate Euclidean distances, suggesting that SpatialNCE does not require a highly precise 3D training signal.
Although it promotes similarity across brains, CytoNet also recovers distinct brain-specific clusters with consistent internal organization.
The embeddings thus resolve consistent internal organization while revealing substantial differences between individuals.
This pattern aligns with interindividual cytoarchitectonic variability demonstrated in probabilistic maps~\citep{Amunts1999} and framed as an essential feature of cortical organization by \citet{Zilles2013a}.
Such variability can exceed inter-area differences~\citep{Amunts1999}.
Brain-specific yet systematically comparable feature spaces provide a basis for studying this variability in relation to functional specialization, development, and disease.

The interaction between SpatialNCE and the neural network architecture may shape this anatomical structure.
Projection heads can discard information that remains useful for downstream tasks~\citep{Chen2020a}.
CytoNet's projection head reduced brain-specific variation in backbone features~(\scref{sub:additional_embedding_analyses}) but did not improve classification, indicating that the suppressed information helps distinguish areas.

Classical profile-based features~\citep{Haug1956,Schleicher1999}, including the structural-prediction baseline used here~\citep{Wagstyl2018}, primarily describe laminar intensity distributions and may oversimplify features such as cell columns.
CytoNet trades the explicit interpretability of classical descriptors for broader coverage.
CytoNet additionally captures cortical curvature and thickness, sectioning effects such as cutting angle, and approximate spatial location.
Individual feature dimensions lack fixed meanings, but the representation can be probed through known anatomical, laminar, and functional variables.
Integrating these sources of variation helps distinguish biological organization from technical influences and supports robust, transferable whole-brain analyses.

Cortical gradients describe cortical organization through continuous low-dimensional axes of cortical shape and structural, functional, or microstructural similarity~\citep{Margulies2016,Wagstyl2018,Paquola2019,Royer2022,Pang2023,Wang2025}.
Both perspectives emphasize that cortical microstructure varies gradually and relates to macroscale functional organization.
However, gradient approaches summarize predefined descriptors into a few interpretable axes, whereas CytoNet learns a high-dimensional representation from local histological images before reduction.
Gradients are therefore one useful lens for interpreting or visualizing CytoNet features, but not the primary output of the model, which is a reusable feature space that preserves local cytoarchitectonic information at cellular resolution.

SpatialNCE imposes a smoothness prior whose scale is set by the bandwidth $\sigma$.
Larger bandwidths bring more distant patches together in feature space and may attenuate sharp boundaries or spatially fragmented areal patterns.
The representation nevertheless recovers broad continuous organization and deviates from this geometry at sharp areal transitions, brain-specific structures, and task-relevant local variations.
The area-classification error-distance and anatomical-$k$ clustering analyses indicate that adjacent and subareal structure survives this smoothness prior.


A pretrained representation shifts the bottleneck of brain mapping from manual annotation to scalable, data-driven analysis, making it feasible to process entire brains and compare across individuals.
CytoNet supports this shift across area classification, cortical layer segmentation, and exploratory data-driven parcellation, with features that transfer beyond the exact training configuration.
Target-specific factors nevertheless remain relevant, since downstream accuracy still differs between brains, and both target-brain exposure during self-supervised pretraining and limited supervised annotations improve adaptation.
A general pretrained model should therefore be refined whenever new brains or acquisition settings introduce stronger domain shifts.

Computational mapping complements rather than replaces expert mapping, which supplies training labels and validation standards.
Experts directly compare adjacent regions but typically annotate selected areas sparsely, whereas CytoNet analyzes isolated patches across complete brains.
Both approaches struggle at heterogeneous areal borders that do not necessarily follow anatomical landmarks.
CytoNet recovered the known two-area subdivisions of the frontal pole and Broca's region across several brains, including one excluded from self-supervised pretraining.
However, internal validity criteria did not consistently identify the number of subdivisions.
This hierarchical complexity suggests combining CytoNet features with dedicated clustering, model-selection, spatial-smoothing, and topological-constraint methods~\citep{Schiffer2021a}.
Such representation-based approaches will become increasingly important as histological datasets grow.

Classification performance dropped for brains excluded from self-supervised pretraining, suggesting that CytoNet benefits from a brain-specific representational \enquote{fingerprint}.
Whether this fingerprint reflects individual cellular architecture, histological batch effects, or other factors remains unclear.
Incorporating another brain requires only approximate alignment of digitized sections to the common reference space, a well-established procedure supported by registration and anchoring tools, and no manual image annotations.
Full CytoNet-R50-ViT (1M) pretraining on nine brains required approximately 1792 GPU-hours, with a ten-brain run estimated at approximately 2000 GPU-hours.
Continued self-supervised pretraining on \brain{9} improved target-brain performance at a substantially lower cost of approximately 133 GPU-hours.
Alternatively, few target-brain annotations can adapt the frozen representation through linear probing.
A practical strategy is therefore to apply the pretrained model directly where appropriate, add limited supervised adaptation when needed, and reserve continued or full pretraining for stronger domain shifts.

Layer segmentation and area classification probe local laminar boundaries and regional cytoarchitectonic organization, respectively.
The same representation serves both scales, and layer segmentation in \brain{20} succeeded without prior self-supervised exposure to that brain.
Because annotations were available only for \brain{20}, this result evaluates pretraining exposure but not cross-brain layer-segmentation performance.
Pretraining thus provides reusable features, while limited target-specific supervision adapts them to new brains or tasks.

Functional networks are strongly organized by cortical location, so structure-function analyses must distinguish microarchitectural information from spatial position.
Against the spatial null, CytoNet features carried parcellation-dependent functional information beyond location, with the strongest evidence for Yeo17.
This suggests that cortical function reflects both local microarchitecture and large-scale anatomical position.
It also motivates integrating histological representations with neuroimaging-derived functional embeddings~\citep{Mesulam1998,Hilgetag2016}.


In summary, CytoNet is a step towards scalable pretrained representations for cellular-resolution brain mapping.
Its main contribution is a feature space that preserves cytoarchitectonic information across whole-brain histological datasets and supports several mapping and interpretation tasks.
Future work should extend this approach to more brains, acquisition protocols, and modalities, and should combine CytoNet features with spatial models that enforce anatomical continuity and topology.
Such developments can help bridge local cellular architecture, individual variability, and systems-level brain organization.


\section{Methods} 
\label{sec:materials_and_methods}

\subsection{Microscopic images of histological human brain sections}\label{sub:microscopic_imaging_of_histological_human_brain_sections} 

The analyses were conducted using \num{6475} cell-body stained histological sections from fourteen postmortem human brains ($462\pm20$ sections/brain, min 433, max 492) from the brain collections of our laboratories in Jülich and Düsseldorf, with a total dataset size of approximately \SI{15}{\tb}.
The processing protocol is detailed in~\citet{Amunts2020} and briefly summarized in the following.

Brains were removed from the skull \numrange{24}{36} hours after death (ethics approval \#4863).
They were chemically fixed with formalin or Bodian, and embedded in paraffin.
Coronal sectioning resulted in \numrange{6000}{7500} histological sections with \SI{20}{\micron} thickness.
Every 15th section (every section for \brain{20}) was mounted on a glass slide and stained for neuronal cell bodies using a modified silver staining~\citep{Merker1983}.
Sections were then digitized using high-throughput light-microscopic scanners (TissueScope HS, Huron Digital Pathology Inc.) at a resolution of \SI{1}{\micron\per\px}.
Resulting images have a median size of $\SI{77000}{\px}\times\SI{105000}{\px}$ (\SI{7.5}{\gb}), with a maximum size of up to $\SI{95000}{\px}\times\SI{136000}{\px}$ (\SI{12}{\gb}).
Brain samples considered in this study have the following numerical identifiers: \brain{1}, \brain{2}, \brain{3}, \brain{4}, \brain{5}, \brain{6}, \brain{7}, \brain{9}, \brain{10}, \brain{11}, \brain{12}, \brain{13}, \brain{14}, and \brain{20}.
Brain sections are referred to by four-digit numbers that increase along the posterior-anterior axis.
Brain \brain{20} refers to the so-called BigBrain model, a 3D-reconstructed dataset based on \num{7404} histological sections~\citep{Amunts2020}.
To be comparable with the non-BigBrain histological datasets, every 15th section of \brain{20} was used in this study.

\subsection{Generation of sampling locations in the cerebral cortex}\label{sub:generation_of_sampling_locations_in_the_cerebral_cortex} 

Sampling locations for contrastive pretraining and downstream feature extraction were prepared by defining locations along the cortical midline in each brain used for the respective analysis.
The cortical midline runs centered between the pial surface and the gray matter surface.
Identifying the cortical midline based on 2D scans of histological sections is sometimes challenging due to the projection of the three-dimensional structure of the cortex onto the 2D image planes.
To address this, each brain was approximately 3D reconstructed by section-to-section alignment, consistent midsurfaces in 3D were computed, and then projected back onto the 2D image planes~\footnote{Code available at \url{https://jugit.fz-juelich.de/inm-1/bda/software/data_processing/brain3d}.}.

Approximate 3D reconstructions of all brains were computed, except for brain \brain{20}, for which the high-resolution anatomical reconstruction of the BigBrain dataset was used.
Reconstructions were created by computing rigid alignments between all pairs of adjacent brain sections~\citep{Dickscheid2019}.
Rigid transformations were estimated from SURF features~\citep{Bay2006a} computed at a resolution of \SI{64}{\micron\per\px}, which were matched using k-nearest neighbor matching and filtered by the RANSAC algorithm~\citep{Fischler1981}.
For a few sections, rigid alignment was not sufficient because these sections were digitized with reversed orientation, producing left–right mirrored images.
Affine transformations for these sections were computed after identifying them in a manual quality check.
Finally, the approximate 3D reconstruction of each brain was computed by aligning all sections to a base section in the center of the section stack of each brain using recursive application of the computed section-to-section transformations.
Limiting the alignment to rigid transformations avoids strong deformations and distortion of the reconstructed brain volumes.

To compute cortical midsurfaces, each section image was segmented~\footnote{Code available at \url{https://jugit.fz-juelich.de/inm-1/bda/software/analysis/tseg}.} into gray matter, white matter, and background (i.e., microscopy slide).
Microscopic scans downscaled to \SI{64}{\micron\per\px} were used for computing this tissue segmentation.
Before segmentation, the contrast of the images was enhanced to better distinguish between gray and white matter.
A minimum filter, a maximum filter, and a mean filter were applied, each with size 5.
In a next step, the contrast was enhanced using contrast limited adaptive histogram equalization (CLAHE, ~\citet{Pizer1987}) with a kernel size of 250, followed by Gaussian blurring (standard deviation 1), followed by another round of minimum, maximum, and mean filters with size 5.
The background class was identified by searching for local minima in the intensity histogram (256 bins) of each image.
Histograms were smoothed using a median filter (size 3) and a mean filter (size 5) to make the process robust against noise.
If more than one minimum was found, the one closest to the Otsu threshold~\citep{Otsu1979} was used.
Pixels identified as tissue using the background segmentation were segmented into gray and white matter using morphological active contours~\citep{Marquez-Neila2014}, a variant of the Chan-Vese segmentation method~\citep{Chan1999}.
Resulting segmentations were cleaned using morphological operations to remove small objects and holes from the mask.
All steps were tuned to prevent tight sulci from being closed during segmentation or cleanup to retain the shape of the cortex.
Obtained segmentation masks were then 3D reconstructed at an isotropic resolution of \SI{300}{\micron\per\vx} using the computed rigid transformations.
For \brain{20}, the tissue segmentation available from the BigBrain dataset was used~\citep{Lewis2014}.

These segmentation volumes were cleaned to remove segmentation errors, imprecise alignment, or histological artifacts.
Tissue defects were detected by smoothing volumes with a median filter of size 3 in the posterior-anterior direction and computing the difference to the input volume.
Larger tissue defects were identified by detecting large connected components in the difference volume, and then replaced by the result of the median filter.
Small parts of detached tissue were removed by extracting all connected components that were smaller than 1\% of the largest tissue component.
Parts of the volume belonging to subcortical gray matter and the cerebellum, which are not handled by the employed segmentation pipeline, were manually identified and excluded using the \emph{3DSlicer} software~\citep{Kikinis2014}.
The manual steps required approximately \SI{30}{\min} per brain.

Following~\citet{Leprince2015}, the Laplacian field in the cerebral cortex was computed using \emph{BrainVisa}~\citep{Riviere2009}.
The Laplacian field approximates the cortical depth, taking the value 0 at the pial surface and linearly increasing to 1 towards the gray-white matter boundary.
The \emph{marching cubes algorithm}~\citep{Lewiner2003} was applied to extract the $0.5$-isosurface from the Laplacian field, which approximates the midsurface through the cortex.
The resulting midsurface meshes were cleaned by removing small isolated connected components, splitting brain hemispheres into separate meshes, fixing topological errors, and computing a Poisson surface reconstruction~\citep{Kazhdan2006} to remove artifacts from reconstruction inaccuracies or segmentation errors.
Isotropic explicit remeshing~\citep{Surazhsky2003} was then applied to remesh all triangle edges to a length of approximately \SI{300}{\micron}.
Mesh processing was performed using the \emph{MeshLab} software~\citep{Cignoni2008}.

The 2D midline through the cortex was derived by projecting the 3D midsurface back onto the 2D images.
For each histological section, the plane that cuts through the midsurface at the location of the respective brain section was determined, and the intersection between this plane and the midsurface, which can be interpreted as \enquote{virtually cutting} the reconstructed brain, was computed.
The intersection was transformed back onto the brain sections by inverting the transformations used for 3D reconstruction.

As a result of the smoothing and cleaning steps in 3D, points transformed from 3D to 2D were not always located exactly in the center of the cortex.
To address this, a refinement step that \enquote{pushes} points towards the cortical midline was applied.
For the refinement, the morphological skeleton of the cortex segmentations was derived.
Laplacian fields between the skeleton and both the pial boundary and gray-white matter boundary were then computed using successive over-relaxation.
Each point was then integrated through the gradient field of the Laplacian fields, limiting the maximum movement to \SI{2}{\mm}.
Points that contained less than 50\% tissue according to the tissue segmentation were excluded from further processing.
In total, \num{6301671} sample points were created ($\num{450119}\pm\num{46879}$ points/brain, min \num{385343}, max \num{539030}).

During pretraining, the presented SpatialNCE loss requires each image patch to be associated with a corresponding spatial location in the brain for computing similarity between samples.
To allow distance computation across brains, it is important for spatial locations to be defined in a common reference coordinate system.
To accomplish this, we made use of the fact that digitized histological sections used for training CytoNet are a subset of the dataset that was used to create the Julich Brain Atlas~\citep{Amunts2020}, for which linear and non-linear transformations from the pixel space of the digitized histological sections and the individual brain template \emph{MNI Colin 27}~\citep{Holmes1998} are available as part of the Julich Brain workflow.
These transformations were used to associate each sampling location in the microscopic images with a corresponding location in the coordinate system of the MNI Colin 27 space.
Details on the used transformation workflow are provided in~\citet{Amunts2020}.
In \scref{sub:extended_scores_for_brain_area_classification}, we also evaluated models pretrained on coordinates from the \emph{MNI 152 ICBM 2009c Nonlinear Asymmetric template space}~\citep{Fonov2011}, obtained by nonlinearly transforming coordinates from MNI Colin 27 space using siibra-python~\citep{Dickscheid2026}; scores are summarized in \scref{tab:cytonet_variants_table}.

To test distances along the folded cortical sheet as a training signal, we trained CytoNet-R50 (200k) + GDist, a geodesic-distance variant of SpatialNCE reported in \scref{tab:cytonet_variants_table}.
We retrieved the left and right Colin27 cortical surface meshes with siibra-python, comprising \num{171259} and \num{170783} vertices, respectively.
Because geodesic distances are too expensive to compute during distributed training, we precomputed all pairwise vertex distances for each hemisphere using the MPI-parallelized \emph{geodist\_mpi} tool\footnote{\url{https://jugit.fz-juelich.de/inm-1/bda/software/data_processing/geodist_mpi}}, which internally uses \emph{potpourri3d}.
The resulting distance matrix required approximately \SI{218}{\gb} of storage.
To reduce small numerical asymmetries introduced by approximate distance computation, each matrix was symmetrized before training.
During pretraining, each sample location in Colin27 RAS coordinates was assigned to the nearest surface vertex, the corresponding row of the precomputed distance matrix was read, and these geodesic distances were passed to the same RBF kernel used by SpatialNCE in place of Euclidean distances.
All other settings, including architecture, pretraining dataset, bandwidth $\sigma$, optimizer, training schedule, augmentations, and downstream area-classification protocol, were kept identical to the Euclidean-distance CytoNet-R50 (200k) baseline.

\subsection{Deep neural network architectures} 
\label{sub:model_architecture}

CytoNet was evaluated with four architecture variants: \emph{R18}, \emph{R50}, \emph{R18-ViT}, and \emph{R50-ViT}.
\emph{R18} and \emph{R50} are modified ResNet18 and ResNet50~\citep{He2016} architectures following~\citet{Schiffer2021}, where the initial image-processing stem is replaced with two convolutional layers ($5\times5$ convolution with stride $4$ and $3\times3$ convolution with stride 1, 64 filters each) and a $2\times2$ maximum pooling operation to account for the significantly larger input image size compared to many other classification tasks.
Each convolutional layer is followed by a batch normalization layer~\citep{Ioffe2015} and ReLU activation.

\emph{R18-ViT} and \emph{R50-ViT} are hybrids between \emph{R18} or \emph{R50} and the \emph{ViT-B} vision transformer architecture~\citep{Dosovitskiy2020}, which is constructed by appending a \emph{ViT-B} vision transformer to the feature map produced by \emph{R18} or \emph{R50}.
The transformer uses learned positional embeddings that are added to the incoming feature maps.
A special class token is prepended to the transformer input sequence, which aggregates information from the entire input image.
See \citet{He2016} and \citet{Dosovitskiy2020} for an in-depth description of the ResNet50 and ViT-B neural network architectures, respectively.

Models trained using these architectures are referred to by a combination of the training paradigm, the model architecture, and the number of pretraining samples (if applicable).
For example, \emph{scratch-R50} refers to a model trained from random initialization using the R50 architecture, \emph{SupCon-R50-ViT} refers to a model trained using supervised contrastive learning with the R50-ViT-B hybrid architecture, and \emph{CytoNet-R18 (1M)} refers to CytoNet pretrained on 1 million samples using the R18 architecture.
We keep the ResNet backbone explicit in all model names to distinguish R18, R50, R18-ViT, and R50-ViT variants.



\subsection{Pretrained foundation model baselines}\label{sub:pretrained_foundation_model_baselines}

We evaluated pretrained foundation model baselines to compare CytoNet with representations from large-scale general-vision~\citep{Oquab2024}, computational-pathology~\citep{Chen2024,Xu2024,Zimmermann2024}, and microscopy segmentation models~\citep{Archit2025}.
For classification tasks, we evaluated DINOv2 (ViT-B/14)~\citep{Oquab2024}, UNI (ViT-L/16)~\citep{Chen2024}, GigaPath (ViT-G/16)~\citep{Xu2024}, and Virchow2 (ViT-H/14)~\citep{Zimmermann2024}.
For cortical layer segmentation, we also evaluated MicroSAM (SAM-ViT-B)~\citep{Archit2025}.
MicroSAM was evaluated only for cortical layer segmentation because it is a segmentation model with a dense image encoder and no native image-level representation for area classification.
All pretrained foundation model baselines were kept frozen, and only task-specific linear prediction heads were trained.
For area-classification tasks, frozen image-level features were extracted tile-wise and mean-aggregated before training a linear classifier.
For cortical layer segmentation, frozen spatial features were extracted tile-wise, assembled on the segmentation output grid, and passed to a pixel-wise linear classifier.
Table~\ref{tab:model_parameter_counts} summarizes the encoder parameter counts for CytoNet and the pretrained foundation model baselines.
Parameter counts are reported for the feature encoder, excluding task-specific classifier heads.

\begin{table}[ht]
	\caption{\textbf{Encoder sizes for CytoNet and pretrained foundation model baselines.}
		Parameter counts are reported for the feature encoder, excluding task-specific classifier heads.
		Models are sorted by parameter count.}
	\label{tab:model_parameter_counts}
	\begin{center}
		\begin{tabularx}{\textwidth}{@{}l l X r@{}}
			\toprule
			Model                           & Encoder   & Pretraining domain        & params. (M) \\
			\midrule
			CytoNet-R18 (ours)              & R18       & cortical cytoarchitecture & 11.21       \\
			CytoNet-R50 (ours)              & R50       & cortical cytoarchitecture & 23.54       \\
			DINOv2~\citep{Oquab2024}        & ViT-B/14  & general vision            & 86.58       \\
			MicroSAM~\citep{Archit2025}     & SAM-ViT-B & microscopy segmentation   & 93.73       \\
			CytoNet-R18-ViT (ours)          & R18-ViT-B & cortical cytoarchitecture & 97.44       \\
			CytoNet-R50-ViT (ours)          & R50-ViT-B & cortical cytoarchitecture & 110.95      \\
			UNI~\citep{Chen2024}            & ViT-L/16  & computational pathology   & 303.35      \\
			Virchow2~\citep{Zimmermann2024} & ViT-H/14  & computational pathology   & 631.24      \\
			GigaPath~\citep{Xu2024}         & ViT-G/16  & computational pathology   & 1134.95     \\
			\bottomrule
		\end{tabularx}
	\end{center}
\end{table}

\subsection{Self-supervised pretraining of CytoNet}\label{sub:self_supervised_pretraining} 

\paragraph*{Dataset.}\label{sub:dataset_for_self_supervised_pretraining} 
Two pretraining datasets were created by randomly sampling \num{200000} and \num{1000000} samples from generated sampling locations in the self-supervised pretraining set~(\cref{sub:generation_of_sampling_locations_in_the_cerebral_cortex}), denoted as \emph{200k} and \emph{1M}, respectively.
No balancing of samples (e.g., to address varying area sizes) was performed.
The self-supervised pretraining set contained \brain{1}, \brain{3}, \brain{4}, \brain{5}, \brain{6}, \brain{7}, \brain{10}, \brain{12}, and \brain{20}.
Brains \brain{2}, \brain{9}, \brain{11}, \brain{13}, and \brain{14} were not used for self-supervised pretraining and were included in feature-space analyses and downstream evaluation.
CytoNet-R50-ViT (1M, no-B20) denotes a matched \brain{20}-exclusion control in which \brain{20} was replaced by \brain{9} during self-supervised pretraining.
This kept the number of pretraining brains and sampled patches unchanged and, for analyses derived from \brain{20}, tested whether performance depended on self-supervised exposure to the target brain.
During training, microscopic image patches were extracted centered at the sampled locations.
Each image patch had a square size of \SI{2048}{\px} at a resolution of \SI{2}{\micron\per\px}, resulting in an effective field of view of approximately \SI{4}{\mm}.
According to \citet{VonEconomo1925}, cortical thickness in the isocortex (before correction for shrinkage from histological processing) varies between \SIrange{3.3}{4.5}{\mm} in the primary motor cortex (Brodmann area 4) and \SIrange{1.9}{2.1}{\mm} in the primary somatosensory cortex (Brodmann area 3).
The field of view is thus sufficiently large to fully capture cytoarchitectonic patterns in most parts of the isocortex.

\paragraph*{Training protocol.} 
\label{sub:training_protocol_cytonet}
CytoNet was trained using Stochastic Gradient Descent (SGD) with Nesterov momentum and a momentum factor of 0.9 in combination with the LARS optimizer~\citep{You2017} with trust coefficient 0.02.
The batch size was $B=2048$, and the learning rate was set to $0.01 \times (B / 256) = 0.08$, which was kept constant over the course of the training.
Weight decay with a factor of 0.0001 was applied to all non-bias parameters of the model.
The temperature parameter $\tau$ for contrastive pretraining was set to 0.07.
Pretraining was performed for 150 epochs.
The RBF kernel for the SpatialNCE loss used a bandwidth of $\sigma=\SI{10}{\mm}$.

Data augmentation was applied to capture typical variations in the data, following the data augmentation strategy detailed in~\citet{Schiffer2021}, which is briefly summarized below.
Image patches were randomly rotated by $\theta \sim U\left[-\pi, +\pi\right]$ ($U[a, b]$: uniform distribution over $[a, b]$), patch center positions were translated in a random direction by $d \sim U\left[0 \textrm{mm}, 0.2\textrm{mm}\right]$, and mirrored vertically with a probability of $50\%$.
Pixel intensities  $x \in \left[0, 1\right]$ were randomly augmented using unbiased gamma augmentation~\citep{Pohlen2017} $\alpha x^\gamma + \beta$ with parameters $\alpha \sim U[0.9, 1.1]$, $\beta \sim U[-0.1, +0.1]$, $\gamma = \frac{\log \left(0.5 + 2^{-0.5} Z\right)}{\log \left(0.5 - 2^{-0.5} Z\right)}$, $Z \sim U[-0.05, +0.05]$.
In addition, images were blurred with an isotropic Gaussian $G_\sigma (x)$ of kernel size $\sigma \sim U[0.125, 1.0]$, or sharpened according to $x + \delta (x - G_{\sigma_u} (x))$ with ${\sigma}_u \sim U[0.125, 1.0]$, $\delta \sim U[0.5, 1.5]$ with probability $25\%$, respectively.
After augmentation, images were standardized to $\left[-1, 1\right]$ before being passed into the model.

Following \citet{Chen2020a}, projection layers were attached to the respective backbone architecture and the contrastive loss was computed on the output of the projection layers.
A fully-connected layer with as many hidden units as the respective backbone output (512 for R18; 2048 for R50; 768 for R18-ViT and R50-ViT), batch normalization~\citep{Ioffe2015}, ReLU, and a final linear layer with $256$ units were applied to the output of global average pooling for \emph{R18} and \emph{R50} or the class token for \emph{R18-ViT} and \emph{R50-ViT}.



\subsection{Surface visualization of CytoNet features}\label{sub:surface_visualization_of_cytonet_features}

For qualitative anatomical visualization of the learned feature space, we extracted backbone features from CytoNet-R50-ViT (1M) at cortical sampling locations from all 14 considered brains~(\scref{fig:pca_on_mesh_comparison}).
Principal component analysis was fitted jointly across these features and the resulting component scores were mapped back to the corresponding cortical sampling locations for each brain.
Surface maps were visualized using a shared color scale per component across brains.
These maps provide low-dimensional summaries of the high-dimensional CytoNet feature space and were interpreted together with quantitative area-classification and clustering analyses.

\subsection{Local visualization of ViT attention and patch-token representations}\label{sub:local_vit_visualization}

To visualize local image regions used by CytoNet-R50-ViT (1M), we extracted self-attention weights from the first transformer layer.
For each displayed image patch, attention weights from the class token to all spatial patch tokens were reshaped to the token grid and overlaid on the corresponding histology image for each attention head.
Attention scores were gamma transformed for visualization only.
To complement attention maps, we visualized spatial patch-token representations using DINO-style principal component analysis~\citep{Caron2021}.
The class token was removed, and the remaining ViT tokens were L2-normalized, whitened, and reshaped to their spatial token grid.
For the supplementary visualization, a three-component PCA was fitted separately for each displayed area across the patch-token embeddings from all displayed samples of that area.
The first three PCA component scores were mapped to red, green, and blue channels after percentile normalization within each area.
Colors are comparable within each area block and should not be compared across different areas.
These patch-token PCA/RGB maps provide a qualitative view of spatially organized token representations.

\subsection{Brain area classification: Julich Brain Atlas}\label{sub:brain_area_classification} 

\paragraph*{Dataset.}\label{sub:datasets_for_brain_area_classification} 
The dataset used to train classifiers for 113 cytoarchitectonic areas was derived from annotations shown in the Julich Brain atlas (version 3.1, both hemispheres,~\citet{Amunts2020}), following the protocol described in~\citet{Schiffer2021}.
The list of areas is provided in \scref{tab:cytoarchitectonic_areas}.
Annotations were available as contours outlining the outer boundaries of each area.
To generate sampling locations for extracting image patches, the cortical midline was first computed from the morphological skeleton of the rasterized contours.
Sampling points were then uniformly spaced along this midline at \SI{1}{\mm} intervals.

From these potential sampling points, datasets for training, testing, and transferability evaluation were created.
Brains were grouped into three categories ---\emph{seen brains}, \emph{transfer brains}, and \emph{unseen brains}--- based on their inclusion in (i) self-supervised pretraining (if applicable) and (ii) supervised training for brain area classification.
The configuration for these datasets is shown in \cref{tab:brain_dataset}.
Throughout the manuscript, we use \emph{unseen brains} for specimens excluded from both self-supervised pretraining and supervised task training.
When an analysis only depends on self-supervised pretraining status, we refer more specifically to brains excluded from self-supervised pretraining.

Sections from \emph{seen brains} were divided into training and test sections, with 80\% used for supervised training and 20\% reserved for testing.
Test sections were selected by choosing every fifth annotated section from each brain.
This setup reflects a realistic use case in which models are applied to unlabeled sections from a brain with existing partial annotations.

\emph{Transfer brains} were included in self-supervised pretraining but excluded from supervised training.
This configuration enables evaluation of how well the learned representations transfer to brains without annotated training data, a common scenario in ongoing brain mapping efforts where histological brain datasets are digitized before they are manually labeled.

\emph{Unseen brains} were excluded from both self-supervised and supervised training.
This strict separation provides a measure of generalization to entirely novel brains.

To address class imbalance in the training set ---caused by variations in the size of cytoarchitectonic areas--- stratified sampling with replacement was applied.
For each area, \num{1200} image patches were sampled, resulting in a balanced training set of \num{135600} patches.
The sampling rate was chosen such that the median ratio between sampled and available patches across all areas was close to 1.
Test datasets were not resampled and reflect the natural distribution of area sizes.

\begin{table}
	\caption{
		\textbf{Configuration for brains used for pretraining, linear probing, and finetuning.}
		For the matched CytoNet-R50-ViT (1M, no-B20) control, \brain{9} was included in self-supervised pretraining while \brain{20} was excluded, and this model is reported separately where indicated.
	}\label{tab:brain_dataset}
	\begin{center}
		\begin{tabular}[c]{c|c|c|c}
			\hline
			dataset        & brain(s)                                                 & pretraining                 & supervised training         \\
			\hline
			\hline
			seen brains    & \brain{1}, \brain{3}, \brain{4}, \brain{5}               & \multirow{2}{*}{\checkmark} & \multirow{2}{*}{\checkmark} \\
			               & \brain{6}, \brain{10}, \brain{12}, \brain{20}            &                             &                             \\
			\hline
			transfer brain & \brain{7}                                                & \checkmark                  & \xmark                      \\
			\hline
			unseen brains  & \brain{2}, \brain{9}, \brain{11}, \brain{13}, \brain{14} & \xmark                      & \xmark                      \\
			\hline
		\end{tabular}
	\end{center}
\end{table}

To assess label efficiency in Julich Brain area classification, we repeated the linear-probing evaluation with target-brain-specific supervised annotation.
The 0\% condition corresponds to the default linear-probing setup described above: for seen brains, the classifier was trained with the regular balanced supervised training set, whereas for unseen brains no labels from the target brain were included.
For non-zero budgets, the linear classifier was trained only on the indicated fraction of available annotations from the respective target brain, with the CytoNet backbone kept frozen.
This evaluates lightweight supervised adaptation to a target brain without full model retraining.
Where indicated, we additionally evaluated CytoNet-R50-ViT (1M, cont. B09), obtained by continuing self-supervised pretraining on \brain{9} for 10 epochs before linear probing.


\paragraph*{Training protocol.} 
\label{sub:training_protocol_cytonet_areas}
Architecture-matched models were trained to classify 113 cytoarchitectonic brain areas by attaching a linear classifier to pretrained CytoNet models and foundation model baselines (\cref{sub:model_architecture,sub:pretrained_foundation_model_baselines}).
Projection layers from pretrained models were discarded and replaced with a single linear layer (without bias) consisting of 113 output units.
This classifier was attached to the global average pooling output (for R50 and R18) or the class token (for R18-ViT and R50-ViT).
Several baseline models were compared to CytoNet:
\begin{itemize}
	\item \textbf{Training from scratch:} full supervised training from randomly initialized weights.
	\item \textbf{Supervised contrastive learning (SupCon)}~\citep{Khosla2020,Schiffer2021}: supervised contrastive pretraining on labeled training samples.
	\item \textbf{SimCLR (200k and 1M)}~\citep{Chen2020a}: self-supervised pretraining from semantic consistency under multi-view augmentation.
	\item \textbf{Pretrained foundation model baselines} Linear probing of pretrained DINOv2~\citep{Oquab2024}, UNI~\citep{Chen2024}, Virchow2~\citep{Zimmermann2024}, GigaPath~\citep{Xu2024} encoders (\cref{sub:pretrained_foundation_model_baselines}).
\end{itemize}
Classifier training for pretrained models (SupCon, SimCLR, CytoNet, and pretrained foundation model baselines) was performed for 30 epochs, while training from scratch was performed for 180 epochs.
Note that the number of epochs is not directly comparable across all models, since training from scratch and SupCon pretraining are limited to annotated samples, while SimCLR and CytoNet make use of unannotated samples as well.
Nevertheless, the chosen number of epochs was sufficient to ensure convergence in both pretraining and supervised training.

We evaluated two training strategies for the SimCLR, SupCon, and CytoNet models:
\begin{itemize}
	\item \textbf{Linear probing:} Only the classifier was trained, while the pretrained backbone and corresponding batch normalization statistics remained frozen. Note that this setting is not applicable for training from scratch.
	\item \textbf{Finetuning:} Both the classifier and backbone weights were optimized jointly.
\end{itemize}

The same terminology was used for scratch models, where only the finetuning setting is applicable, and for pretrained foundation model baselines, where only the frozen-encoder linear-probing setting was evaluated.

For the pretrained foundation model baselines, each \SI{2048}{\px} histology patch was split into $4 \times 4$ non-overlapping \SI{512}{\px} tiles.
Grayscale histology tiles were repeated to three channels, converted from the standardized training range back to $[0, 1]$, and normalized using the respective pretrained model normalization.
DINOv2 tiles were resized to \SI{518}{\px}, UNI, and Virchow2 tiles were resized to \SI{224}{\px}, and GigaPath tiles were resized to \SI{256}{\px} and center-cropped to \SI{224}{\px}.
Tile-level encoder features were averaged across the 16 tiles before the task-specific classifier.
Only this classifier was trained.

All models were trained using categorical cross-entropy loss and the same data augmentation protocol as used during CytoNet pretraining (\cref{sub:training_protocol_cytonet}).

Unless otherwise noted, all models were trained using SGD with Nesterov momentum and a scaled learning rate of $0.01 \times (B / 256) = 0.08$, with batch size $B=2048$, and weight decay of 0.0001 applied to all non-bias parameters.
Model-specific adjustments were necessary to stabilize training and included:
\begin{itemize}
	\item \textbf{scratch-R50-ViT} and \textbf{CytoNet-R50-ViT (200k)}: learning rate reduced to 0.008.
	\item \textbf{SupCon-R50-ViT}: trained with AdamW and learning rate 0.001.
\end{itemize}
The optimal learning rate was identified with a model-specific parameter sweep with values (0.0001, 0.001, 0.01, 0.1), and the best-performing value for each model was used for the final training.

To evaluate the effect of the projection layer, we also trained and tested \emph{CytoNet-R50-ViT (1M) + Proj}, where the classifier was attached to the output of the pretrained projection layer; linear-probe scores are summarized in \scref{tab:cytonet_variants_table}.
Both linear probing and finetuning were performed to assess whether the projected feature space contains sufficient task-relevant information.


\subsection{Brain area classification: Allen Adult Human Brain Atlas}\label{sub:allen_brain_area_classification}

\paragraph*{Dataset.}\label{sub:datasets_for_allen_brain_area_classification} 
The Allen Adult Human Brain Atlas~\citep{Ding2016} provides a cellular-resolution, Nissl-stained coronal series of the adult human brain acquired independently of our own dataset, serving as an external test of generalization across laboratories and acquisition protocols.
A single-donor series of 106 sections was obtained through the Allen SDK~\footnote{\url{https://github.com/alleninstitute/allensdk}}, and downsampled to a resolution of \SI{2}{\micron\per\px}, matching the resolution used for CytoNet pretraining (\cref{sub:dataset_for_self_supervised_pretraining}).
Cortical areas were taken from the anatomical annotations distributed with the atlas: annotation contours were mapped onto the Allen brain ontology, and the label vocabulary was defined as the deepest descendant cortical structures under the \enquote{neocortex} node of the ontology, yielding 95 cortical subdivisions, many named after Brodmann~\citep{Brodmann1909} areas, e.g.\ the primary motor cortex, area~4.

Sampling locations were generated following the same protocol as for the Julich Brain areas (\cref{sub:datasets_for_brain_area_classification}): the cortical midline was computed from the rasterized annotation contours and sampled uniformly at \SI{1}{\mm} intervals, and image patches of \SI{2048}{\px} at \SI{2}{\micron\per\px} were extracted at each location during training.
Sections were split section-wise into 85 training and 21 test sections (80/20, alternating split).
All data originate from a single donor, so this split measures section-wise generalization within that donor; cross-brain generalization is evaluated in the Julich Brain Atlas experiments.
To address class imbalance, stratified sampling with replacement was applied to the training sections, drawing \num{297} patches for each of the 95 areas represented in the training sections (approximately the median number of available sampling points per represented area) for a balanced training set of \num{28215} patches.
The test set was not resampled and reflects the natural distribution of area sizes (\num{9707} patches).

\paragraph*{Training protocol.}\label{sub:training_protocol_allen_areas} 
Models were trained to classify the 95 Allen cortical subdivisions by attaching a single linear layer with 95 output units to the pretrained encoders, following the area-classification protocol described in \cref{sub:training_protocol_cytonet_areas}.
We evaluated a model trained from scratch (R50-ViT), CytoNet-R50-ViT (1M) under both linear probing and finetuning, and the pretrained foundation model baselines DINOv2, UNI, Virchow2, and GigaPath as frozen tiled feature extractors (\cref{sub:pretrained_foundation_model_baselines}), using the same tiling, aggregation, and preprocessing as for the Julich Brain areas.
All models were trained for 100 epochs with batch size $B=2048$ using SGD with Nesterov momentum and the LARS optimizer, weight decay of 0.0001 on all non-bias parameters, and a constant learning rate scaled as $\text{base\_lr} \times (B / 256)$.
The base learning rate was selected per model: 0.1 for CytoNet linear probing, 0.001 for CytoNet finetuning and DINOv2, and 0.01 for UNI, Virchow2, and GigaPath.
The same augmentation protocol as during CytoNet pretraining (\cref{sub:training_protocol_cytonet}) was used.
Performance was reported as macro-averaged F1 score averaged over three repeated training runs with different model initialization seeds.

To assess label efficiency, each model was trained on balanced subsets comprising 1, 5, 10, 20, and 50\% of the training patches, as well as on the full training set.
Subsampling was performed by reducing the number of samples per area to 3, 15, 30, 59, and 149, respectively.
All models were evaluated on the full test set.

\subsection{Cortical layer segmentation}\label{sub:cortical_layer_segmentation} 

\paragraph*{Dataset.}\label{sub:dataset_for_cortical_layer_segmentation} 
For cortical layer segmentation, a dataset of 913 high-resolution microscopic image patches at \SI{1}{\micron\per\px} resolution was used, each manually annotated with segmentation masks for the six cortical layers as well as background (see \cref{fig:results_layer_segmentation}, B).
The dataset extends a publicly available resource described in~\citet{Dickscheid2021e}.
All annotated layer-segmentation patches were taken from \brain{20} (BigBrain).
To match the resolution used during CytoNet and baseline pretraining, images were downsampled to \SI{2}{\micron\per\px}, and segmentation masks were rescaled to $32 \times 32$ pixels (\SI{128}{\micron\per\px}) to match the output resolution of our models.
Irregularly shaped patches were converted to square format using mirror padding to ensure compatibility with the model input size.

The dataset was split into a fixed 80-20 split (732 training and 184 test patches).
To simulate varying annotation budgets, 1\%, 5\%, 10\%, 20\%, and 100\% subsets of the training pool were sampled.
Each training subset was further split into five folds.
For each fold, a model was trained and then evaluated on the fixed test set.
This repeated sampling design allowed estimating the stability of model performance across different training subsets of equal size.
All results are reported on the held-out test set.
Mean and standard deviation of class-wise macro-F1 scores across the five trained models for each configuration are reported.
To ensure a fair evaluation, mirror padding used to make irregularly shaped images compatible with the models was removed before computing scores.
Thus, for CytoNet-R50-ViT (1M), the layer-segmentation train and test patches came from a brain included in self-supervised pretraining, although the held-out test patches were never used for supervised training.
We therefore repeated the layer-segmentation experiment with CytoNet-R50-ViT (1M, no-B20).
This control is particularly informative because \brain{20} is also used for structural-property prediction, so holding it out helps assess how much BigBrain-based downstream analyses depend on prior self-supervised exposure to BigBrain.


\paragraph*{Training protocol.}\label{sub:training_protocol_cortical_layer_segmentation} 
Models were trained to classify the layer structure of the isocortex by attaching a pixel-wise linear classifier to R50 or R50-ViT architectures (\cref{sub:model_architecture}).
Projection layers from pretrained models were discarded and replaced with a $1\times 1$ convolutional layer (without bias) with 7 channels (six layers plus background).
This classifier was attached to the final spatial feature map for R50 models or the reshaped patch-token feature map for R50-ViT models.
Similar to brain area classification (\cref{sub:brain_area_classification}), training from scratch, SimCLR-R50 and SimCLR-R50-ViT variants~\citep{Chen2020a}, CytoNet-R50 and CytoNet-R50-ViT variants, and the pretrained foundation model baselines described in \cref{sub:pretrained_foundation_model_baselines} were compared.
CytoNet, SimCLR, and scratch models were trained using linear probing and finetuning.
The pretrained foundation model baselines were evaluated with frozen encoders and linear probing only.
MicroSAM was used only for cortical layer segmentation, where its dense encoder can be compared through pixel-wise linear probing.
For DINOv2, UNI, Virchow2, GigaPath, and MicroSAM, grayscale input patches were repeated to three channels and normalized with the respective pretrained-model preprocessing.
DINOv2 used non-overlapping \SI{518}{\px} tiles, UNI used \SI{448}{\px} tiles, Virchow2 and GigaPath used \SI{512}{\px} tiles, and MicroSAM used \SI{1024}{\px} tiles.
The tile stride equaled the tile size for all pretrained foundation model baselines.
MicroSAM features were extracted with the frozen \emph{vit\_b\_histopathology} image encoder.
Frozen tile features were assembled on a $32 \times 32$ output grid and passed to a task-specific $1\times1$ convolutional classifier with seven channels.
The linear classifiers for pretrained foundation model baselines were trained for 50 epochs with batch size 1.
CytoNet, SimCLR, and scratch models used either R50 or R50-ViT architectures.
All models were trained with the AdamW optimizer using a learning rate of 0.001, weight decay of 0.01, and categorical cross-entropy loss.
No data augmentation was applied during training.
For layer segmentation, we used a single shared optimization protocol across models to keep the comparison matched; linear probing was the primary comparison for this task, with finetuning reported as a secondary analysis.


\subsection{Predicting structural variations in cytoarchitecture}\label{sub:predicting_structural_variation_in_cytoarchitecture} 

\paragraph*{Extraction of morphological features from BigBrain.}\label{sub:computation_of_gross_anatomical_properties_for_correlation_analysis} 
Morphological features for brain \brain{20} were extracted using the 3D reconstruction of the BigBrain dataset~\citep{Amunts2013}.
Cortical curvature was computed using the \emph{highres-cortex}~\citep{Leprince2015} module of the \emph{BrainVisa} software~\citep{Riviere2009} based on the gray-white matter segmentation available through the \emph{siibra tool suite}~\citep{Dickscheid2026}.
To assess the cutting direction, the angle between the histological cutting plane and the cortical surface normal was measured.
Surface normals were approximated by computing the gradient of a Laplace field defined between the pial surface and the gray–white matter boundary.
The cutting direction was then quantified as the angle between these gradient vectors and the anterior–posterior axis, which corresponds to the histological cutting direction in BigBrain.
Low angles indicate near-orthogonal slicing relative to the cortical sheet, whereas high angles reflect more oblique cuts, which can obscure the cortical lamination pattern~\citep{Schleicher1999}.
Cortical thickness and layer-specific thicknesses (layers I–VI) were obtained from surface meshes described in~\citet{Wagstyl2020}, accessible via the \emph{siibra} suite.
Thickness was computed as the Euclidean distance between corresponding mesh vertices across laminar surfaces and mapped to the closest points on the cortical midline.
Finally, computed features were assigned to points sampled along the cerebral cortex (\cref{sub:generation_of_sampling_locations_in_the_cerebral_cortex}) based on location in BigBrain space.

\paragraph*{Extraction of layer-wise cell-density from microscopic image patches.}\label{sub:layerwise_cell_density_from_microscopic_image_patches} 
Layer-wise average cell densities were computed from the cortical image patches described in~\cref{sub:dataset_for_cortical_layer_segmentation}.
For each patch cell density was estimated using a kernel density estimator with a kernel bandwidth of \SI{100}{\micron}, based on the positions of segmented cell bodies~\citep{Upschulte2022}.
The resulting density maps were then averaged within each cortical layer, yielding one average cell density value per layer and patch.

\paragraph*{BigBrain intensity profiles.}\label{sub:bigbrain_intensity_profiles} 
Intensity profiles~\citep{Wagstyl2022} sampled across the BigBrain dataset~\citep{Amunts2013} capture depth-dependent structural variations by measuring pixel intensity gradients across the cortical sheet.
They were used as a baseline because laminar intensity profiles and profile-derived gradients are widely used descriptors for microstructural organization in BigBrain-based analyses~\citep{Wei2018,Wagstyl2018,Paquola2019}.
A publicly available dataset from~\citet{Wagstyl2022} was used, comprising \num{327684} cortical profiles.
Each profile contains 200 equidistant intensity values sampled along a 1-pixel wide line perpendicular to the cortical surface, extending from the pial boundary to the gray–white matter interface.
Prior to correlation analysis, all intensity values were standardized via z-scoring.

\paragraph*{Correlative analysis of CytoNet features and BigBrain intensity profiles.}\label{sub:correlative_analysis} 
We conducted a correlative analysis between feature embeddings and known morphological and structural properties of the cerebral cortex.
Learned feature representations of CytoNet-R50-ViT (1M) were compared with structural features extracted from brain \brain{20}, including cortical curvature, cortical thickness, layer-specific thickness, cutting direction, and layer-wise cell densities.
CytoNet-R50-ViT (1M) included \brain{20} during self-supervised pretraining.
Where indicated, the same feature extraction, PCA, and regression analysis was repeated with CytoNet-R50-ViT (1M, no-B20), in which \brain{20} was excluded from self-supervised pretraining.
In parallel, we conducted the same analysis using depth-wise intensity profiles (see above) to serve as a baseline for comparison.
Because intensity profiles alone omit cortical geometry, we additionally concatenated the full intensity profiles with cortical curvature, cortical thickness, or both.
We also evaluated a coordinate-only baseline using the BigBrain-space $x$, $y$, and $z$ coordinates of each sampled location.

For both CytoNet features and intensity profiles, correlations against all target features were evaluated using two input feature sets: the raw features, and principal component projections with dimensionalities 1, 16, 32, 64, 128, and 256.
Principal component analysis (PCA) was fitted once on the entire dataset and applied consistently across all folds.
To quantify predictive strength, separate linear regression models for each target feature were trained.
Models were evaluated using five-fold cross-validation, with $R^2$ scores computed on the held-out test folds and reported as mean $\pm$ standard deviation across folds.
All regressions were implemented using \emph{scikit-learn} with default hyperparameters.
Input-target combinations in which a target variable was also included as an input feature were omitted to avoid self-prediction.
$R^2$ scores for the prediction of layer-wise cell density using full-dimensional CytoNet features are not available, as the number of samples in each fold (730) was not sufficient to fit a linear model based on all dimensions of the CytoNet features (768).

To further interpret the latent structure of CytoNet representations, feature attribution was performed for the first 16 principal components of the CytoNet feature space.
Linear attribution weights were computed by fitting a separate linear regressor to each target property and normalizing the regression coefficients by the ratio of input to output standard deviation.
This normalization ensures that attributions are comparable across targets with differing dynamic ranges.

To evaluate the generalizability of spatial encoding in CytoNet representations, regressors were trained on data from brain \brain{20} and applied to spatial locations in other brains (\scref{tab:location_prediction}).
No finetuning or domain adaptation was performed across brains in this analysis.



\subsection{Decoding functional organization from cytoarchitecture}\label{sec:methods_predicting_brain_function}

We evaluated whether CytoNet representations can decode macroscale functional organization on the cortical surface.
CytoNet-R50-ViT (1M) features were computed for BigBrain \brain{20}, assigned to the BigBrain midsurface by nearest-neighbor lookup in 3D coordinate space, and mapped to fs\_LR surface space for all decoding analyses.
CytoNet-R50-ViT (1M), used for functional decoding, included \brain{20} during self-supervised pretraining.
Where indicated, CytoNet-R50-ViT (1M, no-B20) was used as the matched \brain{20}-exclusion control for BigBrain-derived feature analyses.

As functional targets, we used the Yeo7 and Yeo17 network parcellations~\citep{ThomasYeo2011} as well as the DiFuMo atlas with 64 functional modes~\citep{Dadi2020}.
Yeo labels were obtained in BigBrain space from the \emph{BigBrainWarp} toolbox~\citep{Paquola2021}.
DiFuMo64 was fetched in MNI152 space using the \emph{siibra} tool suite~\citep{Dickscheid2026} and mapped to fs\_LR (32k) surface space~\citep{VanEssen2012} with nearest-neighbor resampling using \emph{neuromaps}~\citep{Markello2022}.

All experiments were performed in fs\_LR (32k) surface space after mapping all features and targets to this space.
CytoNet feature maps were warped from BigBrain space to fs\_LR using \emph{BigBrainWarp}.
For baseline comparisons, we also included histological gradients (Hist-G1 and Hist-G2)~\citep{Paquola2019}, microstructural gradients derived from quantitative T1 (qT1; Micro-G1 and Micro-G2)~\citep{Royer2022}, and functional gradients (Func-G1 and Func-G2)~\citep{Margulies2016}, obtained via \emph{BigBrainWarp} and evaluated in the same fs\_LR space.
We also evaluated a spatial baseline using fs\_LR vertex coordinates (x,y,z), and a combined feature set concatenating CytoNet PCA features with coordinates.

To obtain low-dimensional CytoNet variants, principal component analysis (PCA) was fitted once on CytoNet features pooled across both hemispheres in fs\_LR space and subsequently applied to each hemisphere.
We evaluated PCA projections with 1, 2, and 32 components, where PCA-$N$ denotes the first $N$ principal components.
Thus, CytoNet PCA-1 and PCA-2 provide low-dimensional comparisons to G1 and G1+G2 gradient baselines, whereas PCA-32 evaluates the predictive information available in a higher-dimensional CytoNet representation.

Functional labels were predicted using multinomial logistic regression with class balancing, implemented in \emph{scikit-learn}.
All input features were standardized within each training fold (zero mean, unit variance) and the transformation was applied to the held-out fold.

\paragraph*{Spatially grouped cross-validation.}
To reduce score inflation from spatial autocorrelation, we used grouped 5-fold cross-validation with groups defined by spatial surface superblocks.
Superblocks were generated by 3-fold explicit isotropic mesh decimation of the fs\_LR surface to obtain a coarser set of vertices, assigning each original vertex to its nearest remeshed vertex, and then clustering block centroids into 100 parcels per hemisphere using $k$-means.
All vertices belonging to the same superblock were kept in the same fold.

\paragraph*{Spin-based spatial null.}
To test whether performance gains over the coordinate baseline can be explained by spatial smoothness alone, we performed a spin-based spatial permutation test on the functional-parcellation labels in fs\_LR space.
For each hemisphere, labels were rotated on the spherical fs\_LR representation using random rotations and reassigned to vertices by nearest-neighbor matching on the sphere, implemented with \emph{neuromaps}~\citep{Markello2022}.
Rotations were performed independently per hemisphere to preserve hemispheric topology, and the same spatially grouped cross-validation procedure was applied after each spin.
We assessed statistical significance of the observed improvement in macro-F1 between CytoNet PCA--32 and the coordinate-only baseline using a one-sided test,
with $p = (1 + \#\{\Delta_{\mathrm{spin}} \ge \Delta_{\mathrm{obs}}\})/(1 + N_{\mathrm{spin}})$.
Reported deltas were computed from the paired CytoNet and coordinate scores used in the spin test and are therefore reported separately from rounded cross-validation means shown in the figure.
When testing multiple parcellations, $p$-values were corrected using Holm--Bonferroni correction.
Performance was quantified by macro-F1 and reported as mean $\pm$ standard deviation across spatially grouped cross-validation folds.
To reduce runtime, spin tests were conducted in parallel using 1024 CPU cores across 4 compute nodes of JURECA-DC (runtime Yeo7: 18 minutes; Yeo17: 43 minutes; DiFuMo64: 160 minutes).



\subsection{Data-driven exploration of cortical areas}\label{sub:data_driven}

\paragraph*{Dataset.}\label{sub:dataset_data_driven_parcellation}
We assessed whether feature embeddings provide a useful substrate for exploratory subdivision of cortical areas in two regions with established cytoarchitectonic substructure:
the frontal pole areas Fp1/Fp2~\citep{Bludau2014} and Broca's areas BA44/BA45~\citep{Amunts1999}.
The analysis was performed in \brain{6}, \brain{9}, and \brain{20}.
\brain{6} and \brain{20} were included in pretraining, whereas \brain{9} was not.
For each region, candidate points were pre-localized in two complementary ways:
(1) by expert annotations available for the respective brain, and
(2) by maximum-probability-map labels from the Julich Brain Atlas 3.1~\citep{Amunts2020}.
Expert annotations were treated as the primary reference because they provide brain-specific cytoarchitectonic delineations.
MPM-based labels were used as an atlas-derived robustness analysis for an atlas-localized workflow; they were not treated as independent individual-brain ground truth and may reflect registration and spatial-prior effects.
Cortical sampling points were generated along the midline as described in \cref{sub:generation_of_sampling_locations_in_the_cerebral_cortex}, without spatial subsampling.

\paragraph*{Clustering.}\label{sub:clustering_data_driven_parcellation}
We compared clustering based on three representations extracted or assigned at the same cortical anchor points:
CytoNet-R50-ViT (1M) backbone features, Virchow2 features~\citep{Zimmermann2024}, and BigBrain laminar intensity profiles~\citep{Wagstyl2018}.
All representations were standardized and reduced with PCA before clustering.
The main analysis used 60 principal components for each representation; sensitivity analyses with fewer components yielded the same method ranking.
Because bilateral pooling can produce clusters that primarily separate hemispheres, clustering was performed separately within each hemisphere.

We used $k$-means clustering and evaluated two complementary settings.
First, for a controlled comparison across representations, $k$ was set to the known number of anatomical subareas present in the expert annotations or MPM labels.
Second, to evaluate whether the cluster number could be estimated from the data, we searched $k=2,\ldots,\min(20,n/15)$, where $n$ is the number of points for a brain, hemisphere, region, and pre-localized approach.
We compared multiple internal criteria, including silhouette score, Calinski--Harabasz index, Davies--Bouldin index, gap statistic~\citep{Tibshirani2001}, and bootstrap stability.
The gap statistic was computed from $k$-means inertia by comparing the observed log within-cluster dispersion with ten uniform reference datasets sampled over the PCA-space bounding box, using the one-standard-error rule for selecting $k$.
The stability-based estimate was used as the data-driven $k^\ast$, while the full set of criteria was inspected to assess whether a unique cluster count was supported.

Cluster-to-label agreement was quantified using adjusted Rand index, adjusted mutual information, purity, and macro-F1 score.
Macro-F1 score was computed after optimal cluster-to-label matching with the Hungarian algorithm and averages F1 score equally across anatomical labels, making it less sensitive to imbalanced subarea sizes.
For agreement at fixed $k$, results were averaged over ten $k$-means initializations.
Cluster identity was not constrained by anatomical proximity or continuity.

\subsection{Computational setup} 
\label{sub:computational_setup}

CytoNet pretraining (\cref{sub:self_supervised_pretraining}) and linear probing for brain area classification (\cref{sub:brain_area_classification}) were performed on \num{16} compute nodes of the supercomputer \emph{JURECA-DC}~\citep{Thornig2021} at Jülich Supercomputing Centre (JSC, Forschungszentrum Jülich, Jülich, Germany).
Each JURECA-DC compute node was equipped with four NVIDIA A100 GPUs ($4\times6912$ CUDA cores, $4\times432$ tensor cores, $4\times\SI{40}{\gb}$ HBM2e memory), two AMD EPYC 7742 CPUs ($2\times64$ cores at \SI{2.5}{\giga\hertz} with hyperthreading), \SI{512}{\gb} memory, and InfiniBand HDR100 interconnect.
Additional pretraining runs were performed on \num{8} Booster nodes of the JUPITER supercomputer at JSC\footnote{\url{https://apps.fz-juelich.de/jsc/hps/jupiter/configuration.html}}.
Each JUPITER Booster node contained four NVIDIA GH200 Grace-Hopper superchips, each pairing a \num{72}-core NVIDIA Grace CPU with an NVIDIA Hopper GPU with \SI{96}{\gb} HBM3 memory, and four InfiniBand NDR200 links.
Training, inference, and evaluation were implemented using \emph{ATLaS}\footnote{\url{https://jugit.fz-juelich.de/inm-1/bda/software/analysis/atlas/atlas}}, a Python framework that enables large-scale neural network training for high-resolution microscopic image data.

\emph{ATLaS} uses PyTorch~\citep{Paszke2019} for neural network training.
Training was parallelized using distributed data parallel (DDP) training, where each GPU processes a subset of samples in each batch and averages gradients before applying parameter updates.
During contrastive pretraining, features computed by each GPU were gathered to compute pairwise similarities across all samples of a batch.
Statistics computed by batch normalization layers were synchronized across GPUs in each training step.
Automatic Mixed Precision (AMP) was applied to improve training performance by computing certain compute operations with reduced floating point precision.
Gradient checkpointing was used for training R50-ViT models, setting checkpoints after all residual blocks of the convolutional network except for the first one, as well as all transformer layers.
The average GPU memory footprint during contrastive pretraining (32 samples per GPU) was 37.6 GiB for \emph{R50} and 34.4 GiB for \emph{R50-ViT} (with gradient checkpointing).
Pretraining (\cref{sub:training_protocol_cytonet}) on 1 million samples for 150 epochs took approximately 28h (1792 GPU hours, 75 GPU days).
Linear probing or finetuning for brain area classification (\cref{sub:brain_area_classification}) on \num{131250} samples for \num{30} epochs took approximately 1.5 hours (96 GPU hours, 4 GPU days).
For the \brain{9} target-brain exposure control, we added samples from \brain{9} to the original self-supervised pretraining dataset and continued training CytoNet-R50-ViT (1M) from the existing checkpoint for 10 additional epochs on this combined dataset.
All other pretraining settings were kept unchanged.
Assuming approximately equal numbers of samples per brain, adding \brain{9} increased the per-epoch compute by about $1/9$.
Using the full-pretraining runtime as reference, this corresponds to approximately $1792 \times 10/150 \times 10/9 \approx 133$ GPU-hours.

Training of models for cortical layer segmentation was performed on a workstation equipped with an NVIDIA RTX 4090 GPU (\SI{24}{\gb} RAM), Intel Core i9-14900K CPU (24 cores, 32 threads, up to \SI{6}{\giga\hertz}), and \SI{192}{\gb} RAM, taking between 5 minutes and 40 minutes per fold, depending on the fraction of used training samples, the model architecture, and whether linear probing or finetuning was used.
The linear regression of structural variations (\cref{sub:predicting_structural_variation_in_cytoarchitecture}) was performed on the same workstation, taking up to 30 seconds per fold, depending on the dimensionality of the input vectors.
No GPU acceleration was used for linear regression.


\section{Acknowledgement}\label{sec:acknowledgement} 

This project received funding from the European Union’s Horizon 2020 Research and Innovation Programme, grant agreement 101147319 (EBRAINS 2.0 Project), the Helmholtz Association port-folio theme “Supercomputing and Modeling for the Human Brain”, the Helmholtz Association’s Initiative and Networking Fund through the Helmholtz International BigBrain Analytics and Learning Laboratory (HIBALL) under the Helmholtz International Lab grant agreement InterLabs-0015, from HELMHOLTZ IMAGING, a platform of the Helmholtz Information \& Data Science Incubator [X-BRAIN, grant number:  ZT-I-PF-4-061], and from the Deutsche Forschungsgemeinschaft (DFG, German  Research Foundation) under the National Research Data Infrastructure – NFDI 46/1 – 501864659.
The authors gratefully acknowledge computing time on the supercomputer JURECA at Forschungszentrum Jülich under grant no. cjinm16.
This project received access to the JUPITER supercomputer, through the JUPITER Research and Early Access Program (JUREAP). JUPITER is funded by the EuroHPC Joint Undertaking, the German Federal Ministry of Research, Technology and Space, and the Ministry of Culture and Science of the German state of North Rhine-Westphalia.


\section{Author contributions}\label{sec:author_contributions} 

\textbf{CS:} Conceptualization, Data curation, Formal analysis, Funding acquisition, Investigation, Methodology, Project administration, Software, Validation, Visualization, Writing – original draft.
\textbf{ZB:} Methodology, Software, Writing – review \& editing.
\textbf{JK:} Methodology, Software, Validation, Writing – review \& editing.
\textbf{JT:} Data curation, Methodology, Software, Writing – review \& editing.
\textbf{KB:} Formal analysis, Investigation, Validation, Writing – review \& editing.
\textbf{HS:} Methodology, Investigation, Validation, Software, Writing – review \& editing.
\textbf{MB:} Methodology, Resources, Software, Writing – review \& editing.
\textbf{TL:}  Resources, Funding acquisition, Writing – review \& editing.
\textbf{KA:} Conceptualization, Funding acquisition, Methodology, Project administration, Resources, Supervision, Validation, Writing – review \& editing.
\textbf{TD:} Conceptualization, Funding acquisition, Methodology, Project administration, Software, Supervision, Validation, Writing – review \& editing.


\section{Declaration of Interests}\label{sec:declaration_of_interests} 

The authors declare no competing interests.


\section{Declaration of generative AI and AI-assisted technologies in the writing process}\label{sec:declaration_of_generative_ai_and_ai_assisted_technologies_in_the_writing_process} 

During the preparation of this work the authors used ChatGPT (OpenAI, versions 5.0, 5.1, 5.5) and Claude (Anthropic, Opus 4.8), in order to improve the clarity and brevity of the text.
After using these tools, the author(s) reviewed and edited the content as needed and take full responsibility for the content of the published article.


\section{Ethics declaration}\label{sub:ethics_declaration} 

The presented study requires no separate ethical approvals.
All usage in this work is covered by a vote of the ethics committee of the Medical Faculty of the Heinrich Heine University Düsseldorf (\#4863).
Postmortem brains were obtained through body donor programs of the anatomical institutes of the universities of Düsseldorf, Rostock, and Aachen, in accordance with legal and ethical regulations and guidelines.
All body donors have signed a declaration of agreement.

\section{Data and code availability}\label{sub:data_and_code_availability} 

\begin{enumerate}
	\item All source code is publicly available under Apache 2.0 licence.
	      The source code used for model training and evaluation is available at \url{https://jugit.fz-juelich.de/inm-1/bda/software/analysis/atlas/atlas}.
	      A docker container for reproducible execution is provided at \url{https://jugit.fz-juelich.de/inm-1/bda/software/analysis/atlas/atlas_container}.
	      Trained model weights, corresponding configuration files, and feature embeddings are available at~\url{https://jugit.fz-juelich.de/inm-1/bda/software/analysis/cytonet_model_zoo}.
	      Code for approximate 3D reconstructions is available at \url{https://jugit.fz-juelich.de/inm-1/bda/software/data_processing/brain3d}.
	      Code for tissue segmentation is available at \url{https://jugit.fz-juelich.de/inm-1/bda/software/analysis/tseg}.
	\item The full histological image dataset used in this study cannot yet be deposited in a public repository because of ongoing processing, data curation, and setup of suitable technical infrastructure to host the dataset.
	      To request access, contact Katrin Amunts (k.amunts@fz-juelich.de, Forschungszentrum Jülich, Germany) with a summary of the planned study.
	      In addition, representative sections of \brain{20} are accessible via the EBRAINS Knowledge Graph (\url{https://doi.org/10.25493/JWTF-PAB}).
	\item The Allen Adult Human Brain Atlas data used for the external transfer experiment are publicly available from the Allen Institute (\url{https://human.brain-map.org/}).
	      Data access and download were performed using AllenSDK (\url{https://github.com/alleninstitute/allensdk}).
	\item Interactive versions of selected figures are available at \url{https://go.fzj.de/cytonet-interactive}.
\end{enumerate}

\newpage

\begin{refcontext}[sorting=nyt]
	\printbibliography
\end{refcontext}

\newpage
\section{Supplementary material}\label{sub:supplementary_material} 

\subsection{List of cytoarchitectonic areas}\label{sub:list_of_cytoarchitectonic_areas} 

\begin{table}
	\caption{
		\textbf{List of 113 brain areas used for brain area classification.}
		Areas are denoted by the nomenclature of the Julich Brain Atlas~\citep{Amunts2020}, e.g., \emph{hOc1} for \emph{human occipital area 1} or \emph{FG1} for \emph{fusiform gyrus area 1}.
	}\label{tab:cytoarchitectonic_areas}
	\begin{center}
		\begin{tabular}{llllllll}
	\hline
	\multicolumn{8}{l}{\textbf{occipital lobe}}                                                                                 \\
	\hline
	\area{hOc1}   & \area{hOc2}   & \area{hOc3v} & \area{hOc4v} & \area{hOc3d}  & \area{hOc4d}  & \area{hOc4la} & \area{hOc4lp} \\
	\area{hOc5}   & \area{hOc6}   &              &              &               &               &               &               \\
	\hline
	\multicolumn{8}{l}{\textbf{parietal lobe}}                                                                                  \\
	\hline
	\area{Ip1}    & \area{Ip2}    & \area{Ip3}   & \area{Ip4}   & \area{1}      & \area{2}      & \area{3a}     & \area{3b}     \\
	\area{5L}     & \area{5M}     & \area{5Ci}   & \area{7PC}   & \area{7A}     & \area{hIP3}   & \area{PF}     & \area{PFcm}   \\
	\area{PFm}    & \area{PFop}   & \area{PFFt}  & \area{PGa}   & \area{PGp}    & \area{hIP1}   & \area{hIP2}   & \area{hIP4}   \\
	\area{hIP5}   & \area{hIP6}   & \area{hIP7}  & \area{hIP8}  & \area{hPO1}   &               &               &               \\
	\hline
	\multicolumn{8}{l}{\textbf{temporal lobe}}                                                                                  \\
	\hline
	\area{FG1}    & \area{FG2}    & \area{FG3}   & \area{FG4}   & \area{Te 1.0} & \area{Te 1.1} & \area{Te 1.2} & \area{Te 2.1} \\
	\area{Te 2.2} & \area{Te 3}   & \area{STS1}  & \area{STS2}  & \area{TeI}    & \area{TI}     &               &               \\
	\hline
	\multicolumn{8}{l}{\textbf{insula}}                                                                                         \\
	\hline
	\area{Ig1}    & \area{Ig2}    & \area{Id1}   & \area{Ig3}   & \area{Id2}    & \area{Id3}    & \area{Id4}    & \area{Id5}    \\
	\area{Id6}    & \area{Ia7}    & \area{Ia1}   &              &               &               &               &               \\
	\hline
	\multicolumn{8}{l}{\textbf{frontal lobe}}                                                                                   \\
	\hline
	\area{4a}     & \area{4p}     & \area{6d1}   & \area{6d2}   & \area{6d3}    & \area{6v1}    & \area{6v2}    & \area{6r1}    \\
	\area{6mp}    & \area{6ma}    & \area{11a}   & \area{11p}   & \area{13}     & \area{Fo4}    & \area{Fo5}    & \area{Fo6}    \\
	\area{Fo7}    & \area{IFJ1}   & \area{IFJ2}  & \area{IFS1}  & \area{IFS2}   & \area{IFS3}   & \area{IFS4}   & \area{8a}     \\
	\area{8b}     & \area{8c}     & \area{8d}    & \area{SFS1}  & \area{SFS2}   & \area{FMS1}   & \area{MFG1}   & \area{44}     \\
	\area{45}     & \area{Op5}    & \area{Op6}   & \area{Op7}   & \area{Op8}    & \area{Op9}    &               &               \\
	\hline
	\multicolumn{8}{l}{\textbf{limbic lobe}}                                                                                    \\
	\hline
	\area{25a}    & \area{25p}    & \area{s24a}  & \area{s24b}  & \area{s32}    & \area{p24a}   & \area{p24b}   & \area{pv24c}  \\
	\area{pd24cd} & \area{pd24cv} & \area{p32}   &              &               &               &               &               \\
	\hline
\end{tabular}

	\end{center}
\end{table}

\begin{table}
	\caption{
		\textbf{List of areas from the Julich Brain Atlas (version 3.1, \citet{Amunts2020}) used in \cref{fig:umap_overview}.}
		For brevity, pre- and postfixes to the area name are omitted (e.g., \enquote{Area hOc1 (V1, 17, CalcS)} is abbreviated as \enquote{hOc1}).
		Indices of areas in the similarity matrices in \cref{fig:umap_overview} are provided.
	}\label{tab:julich_brain_areas}
	\begin{center}
		\begin{tabular}{r p{0.17\textwidth} r p{0.17\textwidth} r p{0.17\textwidth} r p{0.17\textwidth}}
	\toprule
	\multicolumn{8}{l}{\textbf{occipital lobe}}              \\
	\midrule
	0   & hOc1  & 3   & hOc3v  & 6   & hOc4lp & 9   & hOc6   \\
	1   & hOc2  & 4   & hOc4d  & 7   & hOc4v  &     &        \\
	2   & hOc3d & 5   & hOc4la & 8   & hOc5   &     &        \\
	\midrule
	\multicolumn{8}{l}{\textbf{parietal lobe}}               \\
	\midrule
	10  & 1     & 18  & 7M     & 26  & PFcm   & 34  & hIP3   \\
	11  & 2     & 19  & 7P     & 27  & PFm    & 35  & hIP4   \\
	12  & 3a    & 20  & 7PC    & 28  & PFop   & 36  & hIP5   \\
	13  & 3b    & 21  & Op1    & 29  & PFt    & 37  & hIP6   \\
	14  & 5Ci   & 22  & Op2    & 30  & PGa    & 38  & hIP7   \\
	15  & 5L    & 23  & Op3    & 31  & PGp    & 39  & hIP8   \\
	16  & 5M    & 24  & Op4    & 32  & hIP1   & 40  & hPO1   \\
	17  & 7A    & 25  & PF     & 33  & hIP2   &     &        \\
	\midrule
	\multicolumn{8}{l}{\textbf{temporal lobe}}               \\
	\midrule
	41  & CoS1  & 47  & OTS1   & 53  & TI     & 59  & Te 2.2 \\
	42  & FG1   & 48  & Ph1    & 54  & TPJ    & 60  & Te 3   \\
	43  & FG2   & 49  & Ph2    & 55  & Te 1.0 & 61  & TeI    \\
	44  & FG3   & 50  & Ph3    & 56  & Te 1.1 &     &        \\
	45  & FG4   & 51  & STS1   & 57  & Te 1.2 &     &        \\
	46  & FG5   & 52  & STS2   & 58  & Te 2.1 &     &        \\
	\midrule
	\multicolumn{8}{l}{\textbf{insula}}                      \\
	\midrule
	62  & Ia1   & 66  & Id10   & 70  & Id5    & 74  & Id9    \\
	63  & Ia2   & 67  & Id2    & 71  & Id6    & 75  & Ig1    \\
	64  & Ia3   & 68  & Id3    & 72  & Id7    & 76  & Ig2    \\
	65  & Id1   & 69  & Id4    & 73  & Id8    & 77  & Ig3    \\
	\midrule
	\multicolumn{8}{l}{\textbf{frontal lobe}}                \\
	\midrule
	78  & 44    & 90  & 6v3    & 102 & Fp1    & 114 & Op10   \\
	79  & 45    & 91  & 8d1    & 103 & Fp2    & 115 & Op5    \\
	80  & 4a    & 92  & 8d2    & 104 & IFJ1   & 116 & Op6    \\
	81  & 4p    & 93  & 8v1    & 105 & IFJ2   & 117 & Op7    \\
	82  & 6d1   & 94  & 8v2    & 106 & IFS1   & 118 & Op8    \\
	83  & 6d2   & 95  & Fo1    & 107 & IFS2   & 119 & Op9    \\
	84  & 6d3   & 96  & Fo2    & 108 & IFS3   & 120 & SFG2   \\
	85  & 6ma   & 97  & Fo3    & 109 & IFS4   & 121 & SFG3   \\
	86  & 6mp   & 98  & Fo4    & 110 & MFG1   & 122 & SFG4   \\
	87  & 6r1   & 99  & Fo5    & 111 & MFG2   & 123 & SFS1   \\
	88  & 6v1   & 100 & Fo6    & 112 & MFG4   & 124 & SFS2   \\
	89  & 6v2   & 101 & Fo7    & 113 & MFG5   &     &        \\
	\midrule
	\multicolumn{8}{l}{\textbf{limbic lobe}}                 \\
	\midrule
	125 & 33    & 128 & s32    & 131 & CA3    & 134 & TrS    \\
	126 & EC    & 129 & CA1    & 132 & DG     & 135 & Tu     \\
	127 & p32   & 130 & CA2    & 133 & HATA   & 136 & TuTi   \\
	\bottomrule
\end{tabular}

	\end{center}
\end{table}

\cref{tab:julich_brain_areas} shows the list of areas from the Julich Brain Atlas (version 3.1, \citet{Amunts2020}) used for analyses presented in \cref{fig:umap_overview}.
\cref{tab:cytoarchitectonic_areas} provides the names for the 113 areas used in cytoarchitectonic classification, selected as the subset of atlas annotations available in at least four brains in the classification dataset.


\subsection{Extended embedding analysis}\label{sub:additional_embedding_analyses} 

\begin{figure}
	\begin{center}
		\includegraphics[width=1.0\textwidth]{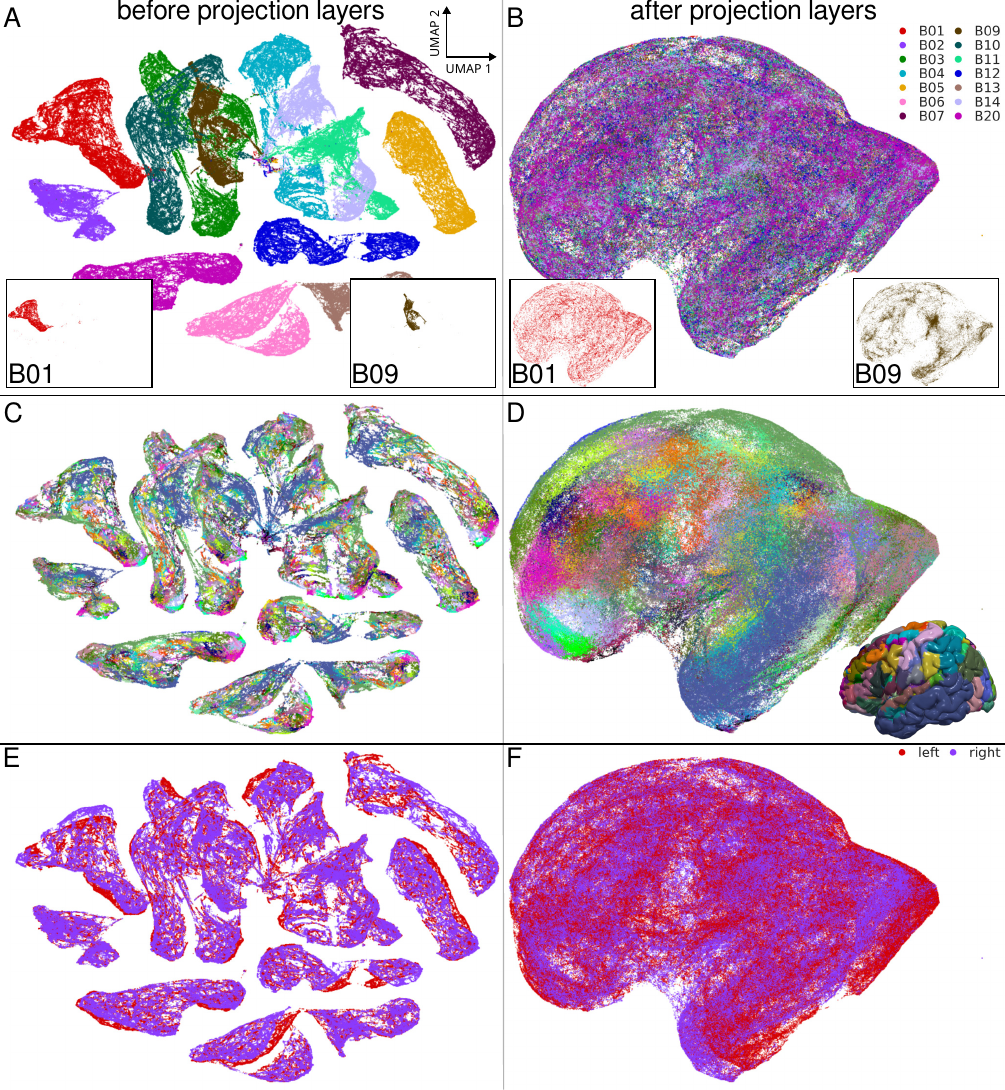}
	\end{center}
	\caption{
		\textbf{2D UMAP embeddings from CytoNet-R50-ViT (1M) backbone features and projection-head outputs.}
		Embeddings are color coded by brain (A, B), Julich Brain labels (C, D), and hemisphere (E, F).
		Backbone features form brain-specific clusters in UMAP space (A) with similar internal arrangement of atlas labels (C) and hemispheres (E).
		Projection-head outputs show stronger cross-brain alignment (B) with consistent atlas-label (D) and hemisphere (F) structure.
		Brains excluded from self-supervised pretraining (e.g., \brain{9}, shown in A+B) appear more compact and less differentiated than pretraining brains in both feature spaces.
	}\label{fig:umap2d_backbone_projection_comparison}
\end{figure}

Following common practice in self-supervised learning~\citep{Chen2020a}, CytoNet was trained with a shallow projection head appended to the backbone.
We compared the UMAP structure of the backbone representation used for downstream analyses and the projection-head output used for the contrastive loss (\cref{fig:umap2d_backbone_projection_comparison}).
Backbone features preserved stronger brain-specific separation, while projection-head outputs showed reduced brain separation and stronger alignment by hemisphere and atlas label.
This is consistent with the role of projection heads in contrastive learning: they shape the space optimized by the pretraining objective and can reorganize information that remains useful in the backbone representation.
Brain separation in the backbone space may reflect a mixture of biological interindividual variability, staining and sectioning differences, registration effects, and model-specific biases.
For brains excluded from self-supervised pretraining (\brain{2}, \brain{9}, \brain{11}, \brain{13}, and \brain{14}), projections retained a global anterior-posterior structure but showed weaker differentiation by atlas label (\chindex{} = 763.02 vs. 3873.23) and hemisphere (\chindex{} = 181.98 vs. 31402.50).

\begin{figure}[tp]
	\begin{center}
		\includegraphics[width=1.0\textwidth]{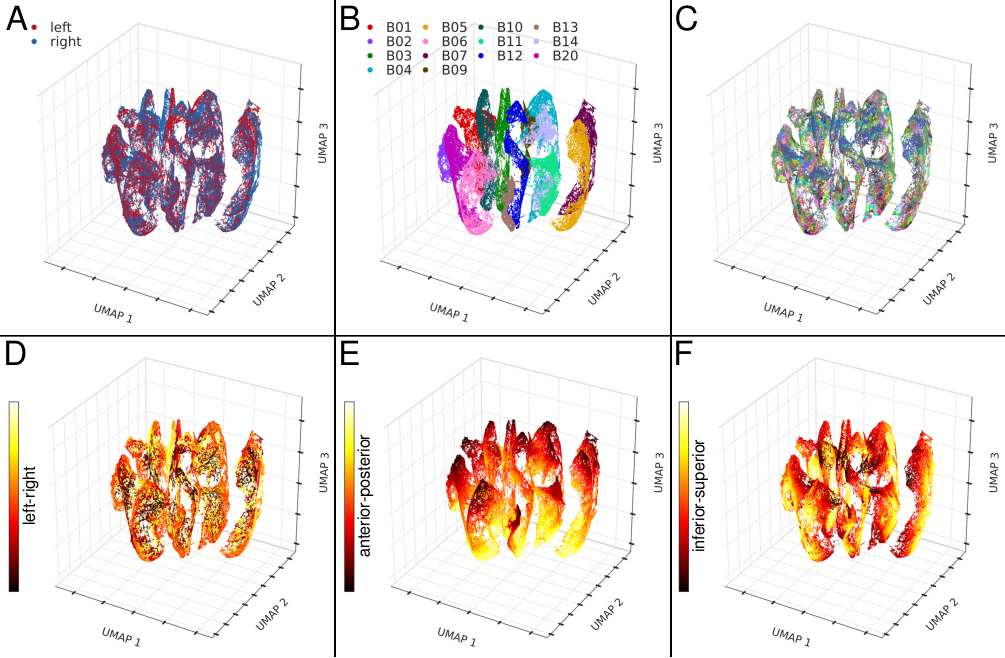}
	\end{center}
	\caption{
		\textbf{3D UMAP plots of CytoNet-R50-ViT (1M) features, colored by hemisphere (A), brain (B), Julich Brain atlas labels (version 3.1,~\citet{Amunts2020}, C), and spatial coordinates in the MNI Colin 27 space (D-F).}
		\textbf{A+D:} Hemispheres manifest as intersecting manifolds per brain.
		\textbf{B:} Brains form distinct clusters.
		\textbf{C:} Brain clusters show consistent internal organization with respect to atlas labels.
		\textbf{D-F:} Coloring by spatial coordinates reveals smooth anatomical gradients within the embedding.
	}\label{fig:umap_3d}
\end{figure}

Complementing our 2D UMAP analyses (\cref{fig:umap_overview}), we visualized 3D UMAP embeddings colored by hemisphere, brain, Julich Brain atlas labels, and spatial coordinates in the MNI Colin 27 space (\cref{fig:umap_3d}).
The embedding separates brains while preserving internal organization with respect to atlas labels.
Coloring the same embedding by MNI coordinates reveals smooth spatial gradients within the brain-specific manifolds, most visibly along anterior-posterior and inferior-superior axes.
UMAP axes are nonlinear embedding coordinates with arbitrary orientation, and these plots provide qualitative visualizations of feature-space organization.
Quantitative spatial encoding is assessed by cross-brain coordinate prediction below.

\begin{figure}[tp]
	\begin{center}
		\includegraphics[width=1.0\textwidth]{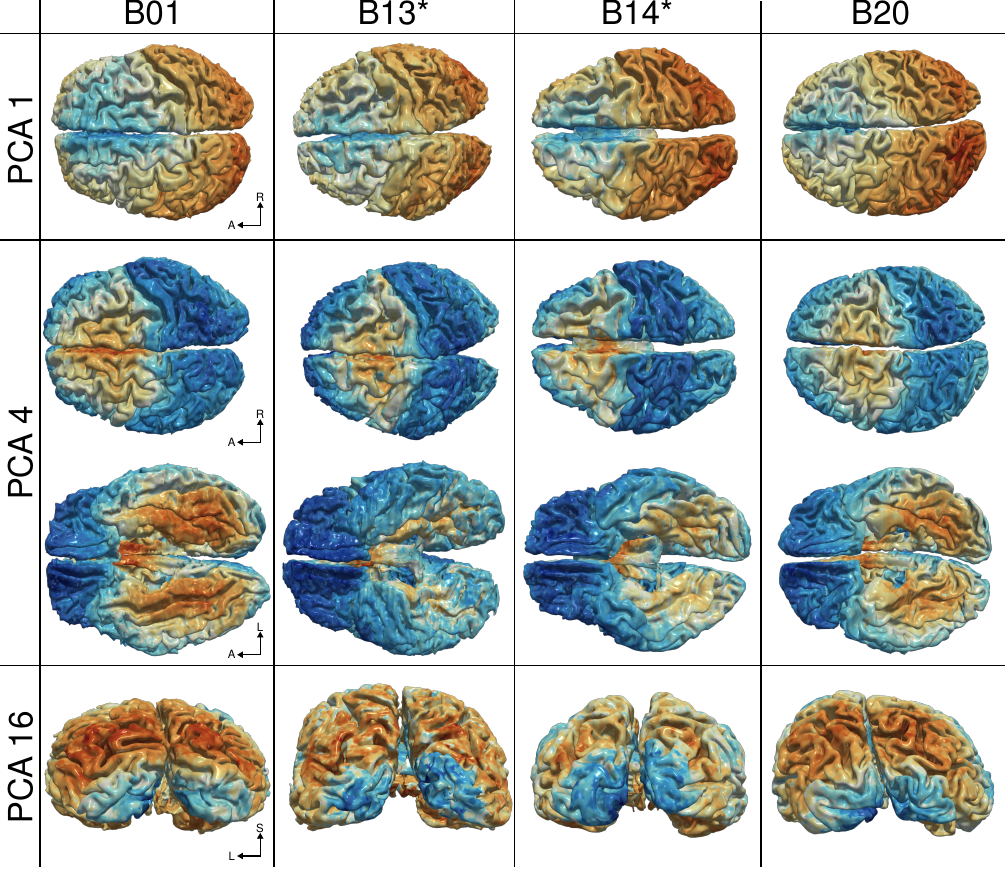}
	\end{center}
	\caption{
		\textbf{Selected PCA components of CytoNet-R50-ViT (1M) features visualized on the cortical surface of brains \brain{1}, \brain{13}, \brain{14}, and \brain{20}.}
		PCA was fitted jointly across all 14 considered brains, and selected components were mapped onto each cortical surface.
		Brains \brain{1} and \brain{20} were included in self-supervised pretraining, while \brain{13} and \brain{14} (marked with $\ast$) were excluded.
		Color scales were normalized per brain to aid visualization.
		Spatial axes: A-P: anterior-posterior; L-R: left-right; I-S: inferior-superior.
	}\label{fig:pca_on_mesh_comparison}
\end{figure}

We visualized selected PCA components of CytoNet-R50-ViT (1M) features on the cortical surface of brains \brain{1}, \brain{13}, \brain{14}, and \brain{20} (\cref{fig:pca_on_mesh_comparison}).
Brains \brain{1} and \brain{20} were included in self-supervised pretraining, while \brain{13} and \brain{14} were excluded.
The surface maps show broad spatial organization that is visible across brains, including brains excluded from pretraining.
PCA components summarize dominant axes of variation in the learned feature space and provide qualitative anatomical visualizations.

The patch-token PCA/RGB maps complement the attention visualizations in the main text by summarizing spatial variation across all patch tokens instead of individual attention heads.
They provide a qualitative view of how CytoNet organizes local token-level structure across cortical areas and across brains included in or excluded from self-supervised pretraining.

\subsection{Cross-brain prediction of spatial coordinates}

To assess how well CytoNet features generalize across brains in terms of spatial encoding, linear regression models were trained to predict MNI coordinates from feature representations.
Specifically, one model was fitted per spatial axis (anterior-posterior, inferior-superior, and left-right) using features extracted from brain \brain{20}.
Models were trained using 5-fold cross-validation on \brain{20}, and then applied to predict spatial coordinates in other brains based on their CytoNet-R50-ViT (1M) features.
Euclidean prediction errors were computed in MNI Colin 27 reference space~(\cref{tab:location_prediction}), where the postmortem brains are being presented after registration~\citep{Amunts2020}.
This approach allows evaluating how consistently CytoNet encodes spatial location across different individuals.
For brains included in self-supervised pretraining, prediction errors were approximately \SI{10}{\milli\metre} in anterior-posterior and inferior-superior directions.
Errors in left-right direction were larger (approximately \SI{20}{\milli\metre}), likely due to structural left-right symmetries.
Brains excluded from self-supervised pretraining (\brain{2}, \brain{9}, \brain{11}, \brain{13}, and \brain{14}) showed higher errors, reaching \SIrange{22}{30}{\milli\metre} in anterior-posterior direction and \SIrange{39}{42}{\milli\metre} in left-right direction.
This indicates reduced spatial generalization without brain-specific pretraining.
Such cross-brain prediction experiments can serve as a proxy for evaluating the general utility of learned features for other structural properties, such as cortical thickness, layer boundaries, or cell density, which are often spatially organized and may benefit from similarly aligned representations.

\begin{figure}
	\begin{center}
		\includegraphics[width=1.0\textwidth]{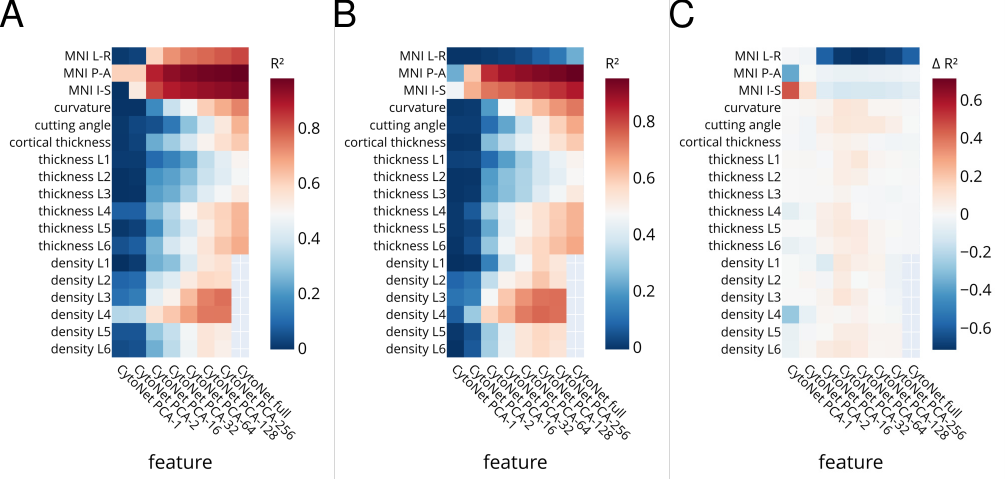}
	\end{center}
	\caption{
		\textbf{Structural-property prediction with and without \brain{20} in self-supervised pretraining.}
		\textbf{A:} Prediction of spatial, morphological, and cytoarchitectonic properties from CytoNet-R50-ViT (1M) features.
		\textbf{B:} Prediction from CytoNet-R50-ViT (1M, no-B20), in which \brain{20} was excluded from self-supervised pretraining.
		\textbf{C:} Difference in $R^2$ between the two models, computed as CytoNet-R50-ViT (1M, no-B20) minus CytoNet-R50-ViT (1M).
		Excluding \brain{20} mostly preserved prediction of non-spatial structural properties, but strongly reduced explicit left-right coordinate prediction in \brain{20}.
	}\label{fig:supplementary_correlation_b20_pretraining_control}
\end{figure}

The \brain{20}-exclusion control in \cref{fig:supplementary_correlation_b20_pretraining_control} tests whether the structural-property prediction results depend on direct self-supervised exposure to \brain{20}.
The comparison shows that most non-spatial structural targets remain predictable after excluding \brain{20} from pretraining, whereas explicit left-right coordinate prediction drops strongly.
This supports the interpretation that \brain{20} exposure mainly affects direct spatial encoding of this axis, while broader cytoarchitectonic and morphological signals remain largely transferable.

\begin{table}
	\caption{
		\textbf{Euclidean distance between true and predicted locations in MNI Colin 27 space.}
		Linear models were fitted on CytoNet-R50-ViT (1M) features from \brain{20} and applied to features from other brains.
		Mean and standard deviation across 5-fold cross validation on \brain{20} are shown.
		For brains included in self-supervised pretraining, prediction errors in anterior-posterior and inferior-superior directions were approximately \SI{10}{\mm}, and approximately \SI{20}{\mm} in left-right direction.
		Brains excluded from self-supervised pretraining (\brain{2}, \brain{9}, \brain{11}, \brain{13}, and \brain{14}) showed higher errors, reaching \SIrange{22}{30}{\mm} in anterior-posterior direction and \SIrange{39}{42}{\mm} in left-right direction.
	}\label{tab:location_prediction}
	\begin{center}
		\begin{tabular}{llll}
\toprule
dimension & left-right & anterior-posterior & inferior-superior \\
 &  &  &  \\
\midrule
B01 & $19.71\pm0.10$ & $11.41\pm0.11$ & $9.79\pm0.13$ \\
B02 & $38.60\pm0.07$ & $26.96\pm0.15$ & $20.34\pm0.05$ \\
B03 & $19.47\pm0.16$ & $11.74\pm0.12$ & $10.91\pm0.12$ \\
B04 & $18.73\pm0.06$ & $11.83\pm0.11$ & $9.95\pm0.07$ \\
B05 & $20.46\pm0.41$ & $12.27\pm0.08$ & $9.66\pm0.07$ \\
B06 & $18.81\pm0.45$ & $13.01\pm0.06$ & $9.61\pm0.07$ \\
B07 & $20.73\pm0.13$ & $15.73\pm0.18$ & $12.05\pm0.28$ \\
B09 & $42.07\pm0.35$ & $22.01\pm0.06$ & $16.34\pm0.08$ \\
B10 & $20.04\pm0.19$ & $12.02\pm0.13$ & $9.69\pm0.04$ \\
B11 & $40.95\pm0.18$ & $30.14\pm0.37$ & $16.84\pm0.18$ \\
B12 & $19.02\pm0.30$ & $10.94\pm0.04$ & $9.63\pm0.10$ \\
B13 & $41.87\pm0.05$ & $25.94\pm0.09$ & $16.44\pm0.05$ \\
B14 & $39.00\pm0.35$ & $22.34\pm0.12$ & $16.05\pm0.12$ \\
\bottomrule
\end{tabular}

	\end{center}
\end{table}

\subsection{Extended scores for brain area classification and layer segmentation}\label{sub:extended_scores_for_brain_area_classification} 

\begingroup
\small
\setlength{\tabcolsep}{3pt}
\begin{longtable}{lllllll}
\caption{
	\textbf{Scores for brain area classification obtained by different models using linear evaluation and finetuning.}
	Mean and standard deviation of scores across three training runs with different random initialization are reported.
}\label{tab:scores_table}\\
\toprule
 & \multicolumn{3}{l}{linear probe} & \multicolumn{3}{l}{finetuning} \\
 & macro-F1 & top-1 acc & top-3 acc & macro-F1 & top-1 acc & top-3 acc \\
\midrule
\endfirsthead
\caption[]{\textbf{Scores for brain area classification obtained by different models using linear evaluation and finetuning} (continued).}\\
\toprule
 & \multicolumn{3}{l}{linear probe} & \multicolumn{3}{l}{finetuning} \\
 & macro-F1 & top-1 acc & top-3 acc & macro-F1 & top-1 acc & top-3 acc \\
\midrule
\endhead
\midrule
\multicolumn{7}{r}{Continued on next page}\\
\endfoot
\bottomrule
\endlastfoot
\multicolumn{7}{l}{\textbf{seen brains}}\\
\midrule
scratch-R50 & - & - & - & $0.39\pm 0.33$ & $0.45\pm 0.32$ & $0.64\pm 0.43$ \\
scratch-R50-ViT & - & - & - & $0.60\pm 0.01$ & $0.65\pm 0.01$ & $0.87\pm 0.01$ \\
SimCLR-R50 (200k) & $0.25\pm 0.00$ & $0.33\pm 0.01$ & $0.56\pm 0.01$ & $0.44\pm 0.18$ & $0.51\pm 0.17$ & $0.77\pm 0.19$ \\
SimCLR-R50 (1M) & $0.15\pm 0.00$ & $0.24\pm 0.01$ & $0.42\pm 0.00$ & $0.47\pm 0.02$ & $0.54\pm 0.02$ & $0.82\pm 0.01$ \\
SimCLR-R50-ViT (200k) & $0.08\pm 0.00$ & $0.19\pm 0.00$ & $0.33\pm 0.00$ & $0.48\pm 0.03$ & $0.54\pm 0.02$ & $0.82\pm 0.02$ \\
SimCLR-R50-ViT (1M) & $0.06\pm 0.01$ & $0.17\pm 0.00$ & $0.31\pm 0.01$ & $0.44\pm 0.08$ & $0.51\pm 0.07$ & $0.78\pm 0.08$ \\
SupCon-R50 & $0.61\pm 0.00$ & $0.66\pm 0.00$ & $0.91\pm 0.00$ & $0.59\pm 0.01$ & $0.65\pm 0.01$ & $0.89\pm 0.01$ \\
SupCon-R50-ViT & $0.61\pm 0.00$ & $0.67\pm 0.00$ & $0.92\pm 0.00$ & $0.51\pm 0.08$ & $0.57\pm 0.08$ & $0.84\pm 0.07$ \\
CytoNet-R18 (200k) & $0.55\pm 0.00$ & $0.62\pm 0.00$ & $0.89\pm 0.00$ & - & - & - \\
CytoNet-R18 (1M) & $0.53\pm 0.00$ & $0.60\pm 0.00$ & $0.88\pm 0.00$ & - & - & - \\
CytoNet-R18-ViT (200k) & $0.50\pm 0.02$ & $0.57\pm 0.01$ & $0.84\pm 0.03$ & - & - & - \\
CytoNet-R18-ViT (1M) & $0.62\pm 0.00$ & $0.67\pm 0.00$ & $0.93\pm 0.00$ & - & - & - \\
CytoNet-R50 (200k) & $0.64\pm 0.00$ & $0.69\pm 0.00$ & $0.93\pm 0.00$ & $0.64\pm 0.01$ & $0.69\pm 0.01$ & $0.92\pm 0.01$ \\
CytoNet-R50 (1M) & $0.54\pm 0.00$ & $0.62\pm 0.00$ & $0.90\pm 0.00$ & $0.67\pm 0.02$ & $0.72\pm 0.01$ & $0.94\pm 0.01$ \\
CytoNet-R50-ViT (200k) & $0.43\pm 0.05$ & $0.52\pm 0.06$ & $0.81\pm 0.06$ & $0.67\pm 0.00$ & $0.72\pm 0.00$ & $0.94\pm 0.00$ \\
CytoNet-R50-ViT (1M) & $0.72\pm 0.00$ & $0.76\pm 0.00$ & $0.97\pm 0.00$ & $0.71\pm 0.02$ & $0.76\pm 0.01$ & $0.95\pm 0.01$ \\
DINOv2 & $0.20\pm 0.05$ & $0.26\pm 0.09$ & $0.48\pm 0.07$ & - & - & - \\
UNI & $0.36\pm 0.00$ & $0.42\pm 0.00$ & $0.68\pm 0.00$ & - & - & - \\
Virchow2 & $0.42\pm 0.00$ & $0.48\pm 0.00$ & $0.74\pm 0.01$ & - & - & - \\
GigaPath & $0.38\pm 0.00$ & $0.44\pm 0.01$ & $0.70\pm 0.01$ & - & - & - \\
\midrule
\multicolumn{7}{l}{\textbf{transfer brain}}\\
\midrule
scratch-R50 & - & - & - & $0.10\pm 0.08$ & $0.19\pm 0.16$ & $0.34\pm 0.26$ \\
scratch-R50-ViT & - & - & - & $0.15\pm 0.02$ & $0.27\pm 0.03$ & $0.49\pm 0.04$ \\
SimCLR-R50 (200k) & $0.11\pm 0.00$ & $0.21\pm 0.01$ & $0.41\pm 0.01$ & $0.14\pm 0.06$ & $0.23\pm 0.11$ & $0.49\pm 0.13$ \\
SimCLR-R50 (1M) & $0.08\pm 0.00$ & $0.20\pm 0.01$ & $0.34\pm 0.01$ & $0.16\pm 0.01$ & $0.29\pm 0.02$ & $0.53\pm 0.02$ \\
SimCLR-R50-ViT (200k) & $0.05\pm 0.00$ & $0.17\pm 0.01$ & $0.28\pm 0.00$ & $0.18\pm 0.02$ & $0.29\pm 0.03$ & $0.57\pm 0.05$ \\
SimCLR-R50-ViT (1M) & $0.05\pm 0.00$ & $0.17\pm 0.01$ & $0.28\pm 0.01$ & $0.15\pm 0.02$ & $0.26\pm 0.03$ & $0.50\pm 0.05$ \\
SupCon-R50 & $0.23\pm 0.01$ & $0.38\pm 0.01$ & $0.67\pm 0.00$ & $0.17\pm 0.02$ & $0.30\pm 0.02$ & $0.55\pm 0.03$ \\
SupCon-R50-ViT & $0.25\pm 0.01$ & $0.40\pm 0.01$ & $0.70\pm 0.01$ & $0.15\pm 0.02$ & $0.28\pm 0.04$ & $0.52\pm 0.06$ \\
CytoNet-R18 (200k) & $0.32\pm 0.00$ & $0.50\pm 0.00$ & $0.80\pm 0.00$ & - & - & - \\
CytoNet-R18 (1M) & $0.29\pm 0.00$ & $0.47\pm 0.00$ & $0.78\pm 0.00$ & - & - & - \\
CytoNet-R18-ViT (200k) & $0.29\pm 0.01$ & $0.45\pm 0.00$ & $0.77\pm 0.00$ & - & - & - \\
CytoNet-R18-ViT (1M) & $0.37\pm 0.00$ & $0.56\pm 0.00$ & $0.88\pm 0.00$ & - & - & - \\
CytoNet-R50 (200k) & $0.33\pm 0.00$ & $0.49\pm 0.00$ & $0.80\pm 0.00$ & $0.18\pm 0.02$ & $0.33\pm 0.02$ & $0.58\pm 0.05$ \\
CytoNet-R50 (1M) & $0.34\pm 0.00$ & $0.52\pm 0.01$ & $0.84\pm 0.00$ & $0.17\pm 0.00$ & $0.33\pm 0.02$ & $0.56\pm 0.02$ \\
CytoNet-R50-ViT (200k) & $0.28\pm 0.04$ & $0.44\pm 0.06$ & $0.73\pm 0.08$ & $0.30\pm 0.01$ & $0.46\pm 0.00$ & $0.77\pm 0.01$ \\
CytoNet-R50-ViT (1M) & $0.38\pm 0.00$ & $0.57\pm 0.00$ & $0.89\pm 0.00$ & $0.26\pm 0.03$ & $0.43\pm 0.03$ & $0.71\pm 0.04$ \\
DINOv2 & $0.05\pm 0.01$ & $0.12\pm 0.03$ & $0.22\pm 0.03$ & - & - & - \\
UNI & $0.09\pm 0.00$ & $0.17\pm 0.00$ & $0.32\pm 0.01$ & - & - & - \\
Virchow2 & $0.10\pm 0.00$ & $0.18\pm 0.00$ & $0.33\pm 0.01$ & - & - & - \\
GigaPath & $0.08\pm 0.00$ & $0.14\pm 0.00$ & $0.26\pm 0.00$ & - & - & - \\
\midrule
\multicolumn{7}{l}{\textbf{unseen brains}}\\
\midrule
scratch-R50 & - & - & - & $0.13\pm 0.11$ & $0.20\pm 0.13$ & $0.36\pm 0.22$ \\
scratch-R50-ViT & - & - & - & $0.16\pm 0.00$ & $0.24\pm 0.00$ & $0.42\pm 0.01$ \\
SimCLR-R50 (200k) & $0.11\pm 0.00$ & $0.18\pm 0.01$ & $0.34\pm 0.01$ & $0.18\pm 0.07$ & $0.25\pm 0.08$ & $0.48\pm 0.12$ \\
SimCLR-R50 (1M) & $0.08\pm 0.00$ & $0.15\pm 0.01$ & $0.28\pm 0.01$ & $0.19\pm 0.03$ & $0.28\pm 0.02$ & $0.51\pm 0.03$ \\
SimCLR-R50-ViT (200k) & $0.04\pm 0.00$ & $0.10\pm 0.00$ & $0.21\pm 0.00$ & $0.20\pm 0.03$ & $0.26\pm 0.02$ & $0.50\pm 0.04$ \\
SimCLR-R50-ViT (1M) & $0.04\pm 0.00$ & $0.11\pm 0.00$ & $0.22\pm 0.01$ & $0.17\pm 0.04$ & $0.24\pm 0.03$ & $0.47\pm 0.06$ \\
SupCon-R50 & $0.26\pm 0.00$ & $0.34\pm 0.00$ & $0.60\pm 0.00$ & $0.20\pm 0.02$ & $0.28\pm 0.01$ & $0.50\pm 0.01$ \\
SupCon-R50-ViT & $0.28\pm 0.00$ & $0.36\pm 0.00$ & $0.63\pm 0.00$ & $0.18\pm 0.04$ & $0.26\pm 0.05$ & $0.48\pm 0.07$ \\
CytoNet-R18 (200k) & $0.29\pm 0.00$ & $0.36\pm 0.00$ & $0.67\pm 0.00$ & - & - & - \\
CytoNet-R18 (1M) & $0.26\pm 0.00$ & $0.33\pm 0.00$ & $0.61\pm 0.00$ & - & - & - \\
CytoNet-R18-ViT (200k) & $0.27\pm 0.02$ & $0.35\pm 0.01$ & $0.64\pm 0.02$ & - & - & - \\
CytoNet-R18-ViT (1M) & $0.32\pm 0.00$ & $0.39\pm 0.00$ & $0.70\pm 0.00$ & - & - & - \\
CytoNet-R50 (200k) & $0.32\pm 0.00$ & $0.40\pm 0.00$ & $0.69\pm 0.01$ & $0.20\pm 0.01$ & $0.30\pm 0.01$ & $0.52\pm 0.02$ \\
CytoNet-R50 (1M) & $0.29\pm 0.00$ & $0.37\pm 0.00$ & $0.66\pm 0.00$ & $0.21\pm 0.02$ & $0.30\pm 0.02$ & $0.52\pm 0.03$ \\
CytoNet-R50-ViT (200k) & $0.24\pm 0.03$ & $0.32\pm 0.05$ & $0.59\pm 0.07$ & $0.28\pm 0.01$ & $0.36\pm 0.00$ & $0.63\pm 0.01$ \\
CytoNet-R50-ViT (1M) & $0.30\pm 0.00$ & $0.37\pm 0.00$ & $0.65\pm 0.00$ & $0.24\pm 0.01$ & $0.34\pm 0.01$ & $0.57\pm 0.01$ \\
DINOv2 & $0.06\pm 0.01$ & $0.10\pm 0.02$ & $0.23\pm 0.02$ & - & - & - \\
UNI & $0.10\pm 0.00$ & $0.15\pm 0.00$ & $0.30\pm 0.00$ & - & - & - \\
Virchow2 & $0.12\pm 0.00$ & $0.18\pm 0.00$ & $0.35\pm 0.00$ & - & - & - \\
GigaPath & $0.11\pm 0.00$ & $0.16\pm 0.01$ & $0.32\pm 0.01$ & - & - & - \\
\end{longtable}
\endgroup

\cref{tab:scores_table} shows macro-F1 scores, top-1 and top-3 accuracy achieved by different models in brain area classification (\cref{sub:brain_area_classification}) on seen, transfer, and unseen subjects.
Performance for finetuning (trainable encoder) and linear probing (frozen encoder) are reported.
The table includes pretrained foundation-model baselines and the unseen-brain evaluation reported in the Results.

\begin{table}
	\caption{
		\textbf{Linear-probe macro-F1 scores for classifying 113 cytoarchitectonic areas from the Julich Brain Atlas with CytoNet variants.}
		GDist denotes a geodesic-distance SpatialNCE variant of CytoNet-R50 (200k).
		Proj denotes CytoNet-R50-ViT (1M) evaluated on projection-head outputs.
		MNI152 denotes CytoNet-R50-ViT (1M) pretrained with coordinates transformed to the MNI 152 ICBM 2009c Nonlinear Asymmetric template space.
		$\Delta$ values are computed relative to the respective CytoNet baseline shown directly above each variant.
		Mean and standard deviation across three training runs with different random initialization are reported.
	}\label{tab:cytonet_variants_table}
	\begin{center}
		\begin{tabular}{lccc}
	\toprule
	model                & seen brains      & transfer brain   & unseen brains    \\
	\midrule
	CytoNet-R50 (200k)   & $0.636\pm 0.001$ & $0.326\pm 0.003$ & $0.322\pm 0.003$ \\
	+ GDist              & $0.586\pm 0.002$ & $0.302\pm 0.004$ & $0.306\pm 0.002$ \\
	$\Delta$             & $-0.050$         & $-0.024$         & $-0.015$         \\
	\midrule
	CytoNet-R50-ViT (1M) & $0.717\pm 0.001$ & $0.385\pm 0.001$ & $0.298\pm 0.001$ \\
	+ Proj               & $0.378\pm 0.000$ & $0.227\pm 0.001$ & $0.217\pm 0.000$ \\
	$\Delta$             & $-0.339$         & $-0.157$         & $-0.081$         \\
	\midrule
	+ MNI152             & $0.578\pm 0.001$ & $0.396\pm 0.003$ & $0.309\pm 0.003$ \\
	$\Delta$             & $-0.138$         & $+0.012$         & $+0.011$         \\
	\bottomrule
\end{tabular}

	\end{center}
\end{table}

\cref{tab:cytonet_variants_table} summarizes targeted controls for distance definition, feature readout, and template space using linear probing.
The geodesic-distance variant of CytoNet-R50 (200k) performed below its matched Euclidean-distance baseline, with macro-F1 differences of $-0.050$ on seen brains, $-0.024$ on the transfer brain, and $-0.015$ on unseen brains.
Evaluating CytoNet-R50-ViT (1M) on projection-head outputs reduced linear-probe performance, indicating that the backbone representation retains more information useful for brain area classification.
Pretraining CytoNet-R50-ViT (1M) with coordinates transformed to MNI152 produced mixed changes relative to the Colin27-based model: seen-brain performance decreased substantially ($\Delta=-0.138$), while transfer- and unseen-brain scores increased slightly ($\Delta=+0.012$ and $\Delta=+0.011$).
This suggests that template choice can affect the learned representation, although the direction of the effect depends on the evaluation split.
Coordinates were non-linearly transformed from MNI Colin 27 to ICBM 152 space using siibra-python~\citep{Dickscheid2026}; natively aligning histological sections to ICBM 152 would provide a more direct assessment of template choice but is challenging due to the limited availability of well-defined landmarks for registration.

\begin{table}
	\caption{
		\textbf{Allen Adult Human Brain Atlas label-efficiency scores.}
		Macro-F1 scores are reported for Allen cortical subdivision classification using balanced subsets of the training data and the full training set.
		Values show mean and standard deviation across three training runs with different initialization seeds.
		Scratch R50-ViT was trained from random initialization, CytoNet-R50-ViT (1M) + FT denotes finetuning, and the remaining pretrained models were evaluated as frozen feature extractors with a linear classifier.
		This table provides the numerical values corresponding to the Allen label-efficiency analysis shown in the main Results figure.
	}\label{tab:allen_label_efficiency_scores}
	\begin{center}
		\begingroup
\small
\setlength{\tabcolsep}{3pt}
\begin{tabular}{lcccccc}
\toprule
 & 1\% & 5\% & 10\% & 20\% & 50\% & 100\% \\
Model &  &  &  &  &  &  \\
\midrule
scratch R50-ViT & $0.07\pm 0.00$ & $0.15\pm 0.00$ & $0.24\pm 0.00$ & $0.30\pm 0.01$ & $0.37\pm 0.01$ & $0.40\pm 0.02$ \\
DINOv2 & $0.11\pm 0.00$ & $0.17\pm 0.00$ & $0.18\pm 0.00$ & $0.19\pm 0.00$ & $0.20\pm 0.00$ & $0.20\pm 0.00$ \\
UNI & $0.15\pm 0.00$ & $0.29\pm 0.00$ & $0.34\pm 0.00$ & $0.36\pm 0.00$ & $0.39\pm 0.01$ & $0.40\pm 0.00$ \\
GigaPath & $0.15\pm 0.00$ & $0.29\pm 0.00$ & $0.35\pm 0.01$ & $0.37\pm 0.00$ & $0.40\pm 0.01$ & $0.40\pm 0.00$ \\
Virchow2 & $0.18\pm 0.00$ & $0.35\pm 0.00$ & $0.43\pm 0.01$ & $0.46\pm 0.00$ & $0.49\pm 0.00$ & $0.49\pm 0.01$ \\
CytoNet-R50-ViT (1M) & $0.17\pm 0.00$ & $0.30\pm 0.00$ & $0.37\pm 0.00$ & $0.42\pm 0.00$ & $0.45\pm 0.00$ & $0.46\pm 0.00$ \\
CytoNet-R50-ViT (1M) + FT & $0.26\pm 0.01$ & $0.46\pm 0.00$ & $0.54\pm 0.01$ & $0.59\pm 0.01$ & $0.63\pm 0.01$ & $0.65\pm 0.00$ \\
\bottomrule
\end{tabular}
\endgroup

	\end{center}
\end{table}

\begin{table}
	\caption{
		\textbf{Scores for Julich Brain Atlas area~\citep{Amunts2020} classification obtained by linear probing of CytoNet-R50-ViT (1M) backbone with cross-validation across transfer brains.}
		Models were pretrained in different training settings, each considering one of the brains \brain{1}, \brain{3}, \brain{4}, \brain{5}, \brain{6}, \brain{7}, \brain{10}, \brain{12}, and \brain{20} as transfer brain, and the remaining as seen brains.
		In all cases, \brain{2}, \brain{9}, \brain{11}, \brain{13}, and \brain{14} were considered as unseen brains.
		Performance on seen and unseen brains is largely independent of the brain used as transfer brain.
		Performance on the respective transfer brain varies slightly, which is likely a result of different subsets of areas that were annotated in each brain.
	}
	\label{tab:performance_metrics_cross_validation}
	\begin{center}
		\begin{tabular}{llll}
\toprule
 & macro-F1 & top-1 acc & top-3 acc \\
\midrule
seen brains & $0.72\pm 0.00$ & $0.75\pm 0.00$ & $0.97\pm 0.00$ \\
transfer brain & $0.36\pm 0.08$ & $0.57\pm 0.02$ & $0.88\pm 0.01$ \\
unseen brains & $0.29\pm 0.00$ & $0.36\pm 0.00$ & $0.64\pm 0.01$ \\
\bottomrule
\end{tabular}

	\end{center}
\end{table}

We evaluated the impact of the selected transfer brain by linear probing of CytoNet-R50-ViT (1M) with training data from different sets of brains~(\cref{tab:performance_metrics_cross_validation}), keeping \brain{2}, \brain{9}, \brain{11}, \brain{13}, and \brain{14} fixed as unseen brains.
Performance for seen and unseen brains was comparable across all choices of transfer brain.
Macro-F1 scores for the transfer brain varied somewhat, reflecting differences in the availability and composition of annotated areas across brains.
Since macro-F1 is particularly sensitive to missing labels (e.g., a missing label that is incorrectly predicted only once contributes a zero to the average), such variability is expected and does not affect the overall conclusion: our experiments with transfer brain \brain{7} are representative for the proposed approach.


\begin{table}
	\caption{
		\textbf{Macro-F1 scores for cortical layer segmentation across different models and training fractions.}
		Mean and standard deviation are reported over 5-fold cross-validation for each model, using either linear probing or finetuning.
		Models were trained on increasing fractions of the training set, and evaluated on a dedicated test set comprising 184 samples.
		At the smallest annotation budgets, finetuning transformer-backbone models produced uniformly low scores in some configurations (scratch-R50-ViT at 1\%: 4/5 folds below 0.10 macro-F1; SimCLR-R50-ViT at 1\%: 3/5 folds; CytoNet-R50-ViT (1M) at 1\%: 5/5 folds).
		At larger budgets, some folds diverged while others reached high scores, most clearly for scratch-R50-ViT at 20\% and 100\% (3/5 and 3/5 folds below 0.10 macro-F1) and CytoNet-R50-ViT (1M) at 100\% (4/5 folds below 0.10 macro-F1; remaining fold 0.81); CytoNet-R50-ViT (1M) at 5\% had 2/5 low-score folds.
		At 100\%, SimCLR-R50-ViT had two folds below 0.30 macro-F1 and CytoNet-R50-ViT (1M, no-B20) had one fold below 0.30 macro-F1, explaining their larger standard deviations.
		The low finetuning scores of CytoNet-R50-ViT (1M) compared with CytoNet-R50-ViT (1M, no-B20) are attributed to greater instability when finetuning the large R50-ViT model on small layer-segmentation datasets.
		All folds are included in the reported means and standard deviations.
	}\label{tab:layer_segmentation_scores}
	\begin{center}
		\begingroup
\small
\setlength{\tabcolsep}{3pt}
\begin{tabular}{llllll}
\toprule
\textbf{linear probe} & 1\% (n=7) & 5\% (n=36) & 10\% (n=73) & 20\% (n=146) & 100\% (n=732) \\
\midrule
SimCLR-R50 (200k) & $0.49\pm 0.02$ & $0.61\pm 0.01$ & $0.66\pm 0.01$ & $0.67\pm 0.02$ & $0.71\pm 0.01$ \\
SimCLR-R50-ViT (1M) & $0.35\pm 0.03$ & $0.50\pm 0.01$ & $0.55\pm 0.01$ & $0.56\pm 0.01$ & $0.59\pm 0.00$ \\
Virchow2 & $0.39\pm 0.00$ & $0.49\pm 0.04$ & $0.55\pm 0.02$ & $0.56\pm 0.02$ & $0.61\pm 0.03$ \\
UNI & $0.63\pm 0.02$ & $0.72\pm 0.01$ & $0.75\pm 0.00$ & $0.77\pm 0.00$ & $0.79\pm 0.01$ \\
DINOv2 & $0.50\pm 0.02$ & $0.61\pm 0.01$ & $0.64\pm 0.01$ & $0.67\pm 0.01$ & $0.71\pm 0.01$ \\
GigaPath & $0.62\pm 0.03$ & $0.71\pm 0.01$ & $0.75\pm 0.01$ & $0.77\pm 0.01$ & $0.79\pm 0.01$ \\
MicroSAM & $0.15\pm 0.01$ & $0.27\pm 0.01$ & $0.36\pm 0.01$ & $0.43\pm 0.00$ & $0.54\pm 0.01$ \\
CytoNet-R50 (200k) & $0.59\pm 0.06$ & $0.72\pm 0.01$ & $0.74\pm 0.00$ & $0.75\pm 0.00$ & $0.77\pm 0.00$ \\
CytoNet-R50-ViT (1M) & $0.63\pm 0.01$ & $0.73\pm 0.00$ & $0.74\pm 0.00$ & $0.75\pm 0.00$ & $0.77\pm 0.00$ \\
CytoNet-R50-ViT (1M, no-B20) & $0.59\pm 0.03$ & $0.71\pm 0.00$ & $0.74\pm 0.00$ & $0.76\pm 0.00$ & $0.78\pm 0.00$ \\
\toprule
\textbf{finetune} &  &  &  &  &  \\
\midrule
scratch-R50 & $0.15\pm 0.10$ & $0.40\pm 0.22$ & $0.65\pm 0.02$ & $0.72\pm 0.02$ & $0.78\pm 0.00$ \\
scratch-R50-ViT & $0.08\pm 0.05$ & $0.52\pm 0.10$ & $0.65\pm 0.05$ & $0.34\pm 0.41$ & $0.20\pm 0.19$ \\
SimCLR-R50 (200k) & $0.15\pm 0.07$ & $0.45\pm 0.05$ & $0.53\pm 0.07$ & $0.69\pm 0.02$ & $0.76\pm 0.03$ \\
SimCLR-R50-ViT (1M) & $0.11\pm 0.03$ & $0.38\pm 0.06$ & $0.50\pm 0.14$ & $0.56\pm 0.19$ & $0.45\pm 0.26$ \\
CytoNet-R50 (200k) & $0.23\pm 0.11$ & $0.57\pm 0.02$ & $0.64\pm 0.05$ & $0.73\pm 0.01$ & $0.78\pm 0.02$ \\
CytoNet-R50-ViT (1M) & $0.05\pm 0.03$ & $0.25\pm 0.18$ & $0.52\pm 0.14$ & $0.78\pm 0.02$ & $0.21\pm 0.33$ \\
CytoNet-R50-ViT (1M, no-B20) & $0.65\pm 0.04$ & $0.78\pm 0.01$ & $0.79\pm 0.01$ & $0.81\pm 0.00$ & $0.72\pm 0.26$ \\
\bottomrule
\end{tabular}
\endgroup

	\end{center}
\end{table}
The CytoNet-R50-ViT (1M, no-B20) rows report the matched \brain{20}-exclusion control.


\subsection{Visualization of data-driven mapping through clustering}\label{sub:visualization_of_data_driven_area_classification_through_clustering} 

\begin{figure}
	\begin{center}
		\includegraphics[width=0.95\textwidth]{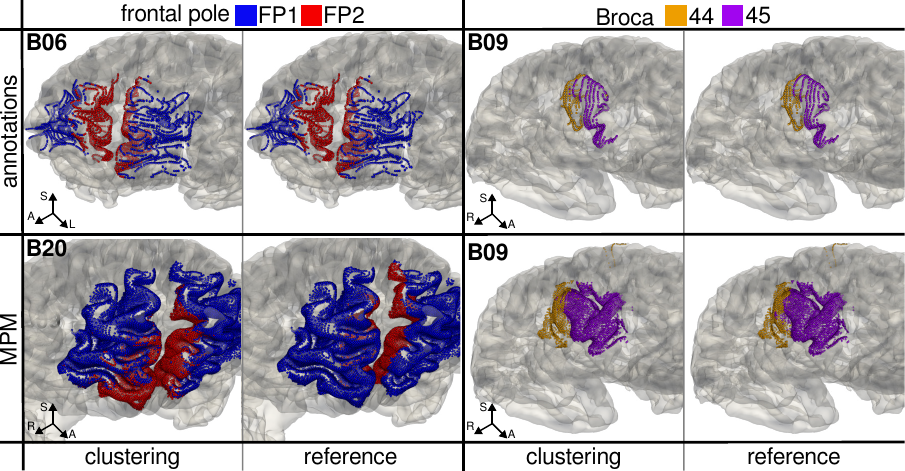}
	\end{center}
	\caption{
		\textbf{Representative visualizations of clustering results for the frontal pole (FP,~\citet{Bludau2014}) and Broca's regions (Broca,~\citet{Amunts1999}).}
		Examples show K-means cluster assignments and corresponding reference labels for annotation-based pre-localization (top row) and Julich Brain Atlas maximum probability map pre-localization (bottom row).
		Clustering was performed separately for each brain, hemisphere, region, and pre-localization strategy; colors denote cluster assignments or reference subdivisions and were manually synchronized between paired visualizations for readability.
	}
	\label{fig:clustering_overview}
\end{figure}

\cref{fig:clustering_overview} provides representative surface examples for the clustering analysis summarized in the main Results figure.
The paired cluster and reference-label maps show where feature-space clusters recover known frontal-pole or Broca subdivisions and where atlas-based pre-localization introduces additional spatial uncertainty.
These examples are intended as qualitative complements to the macro-F1 scores reported in the main figure.

\begin{figure}
	\begin{center}
		\includegraphics[width=0.95\textwidth]{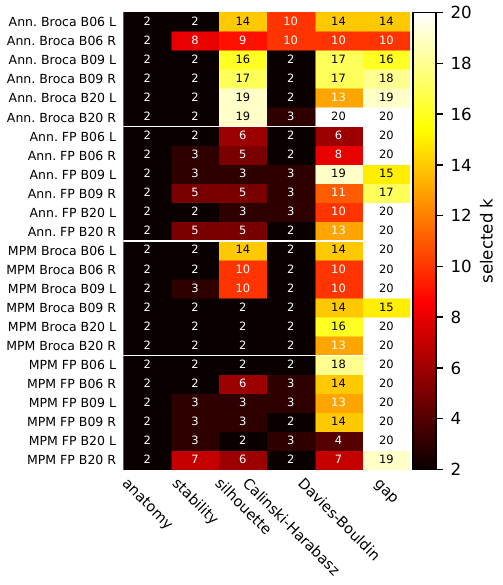}
	\end{center}
	\caption{
		\textbf{Data-driven cluster-number criteria for frontal pole~\citep{Bludau2014} and Broca's region~\citep{Amunts1999} clustering.}
		Heatmap entries show the number of clusters selected by each internal criterion for CytoNet-R50-ViT (1M) features, separately for each region, brain, hemisphere, and pre-localization strategy.
		The criteria do not converge on a unique cluster count across cases.
	}
	\label{fig:clustering_k_selection_summary}
\end{figure}

To assess whether the number of subdivisions could be selected directly from the data, we compared several internal criteria across candidate cluster numbers (\cref{fig:clustering_k_selection_summary}).
Stability denotes bootstrap stability, i.e., the reproducibility of clustering assignments under resampled data.
The criteria did not consistently select the anatomical number of subdivisions across brains, hemispheres, regions, and pre-localization strategies, supporting our interpretation that automatic cluster-number discovery remains a separate model-selection challenge.

\subsection{Investigation of shortcut learning in SimCLR}\label{sub:investigation_of_shortcut_learning_in_simclr} 

\begin{figure}
	\begin{center}
		\includegraphics[width=0.95\textwidth]{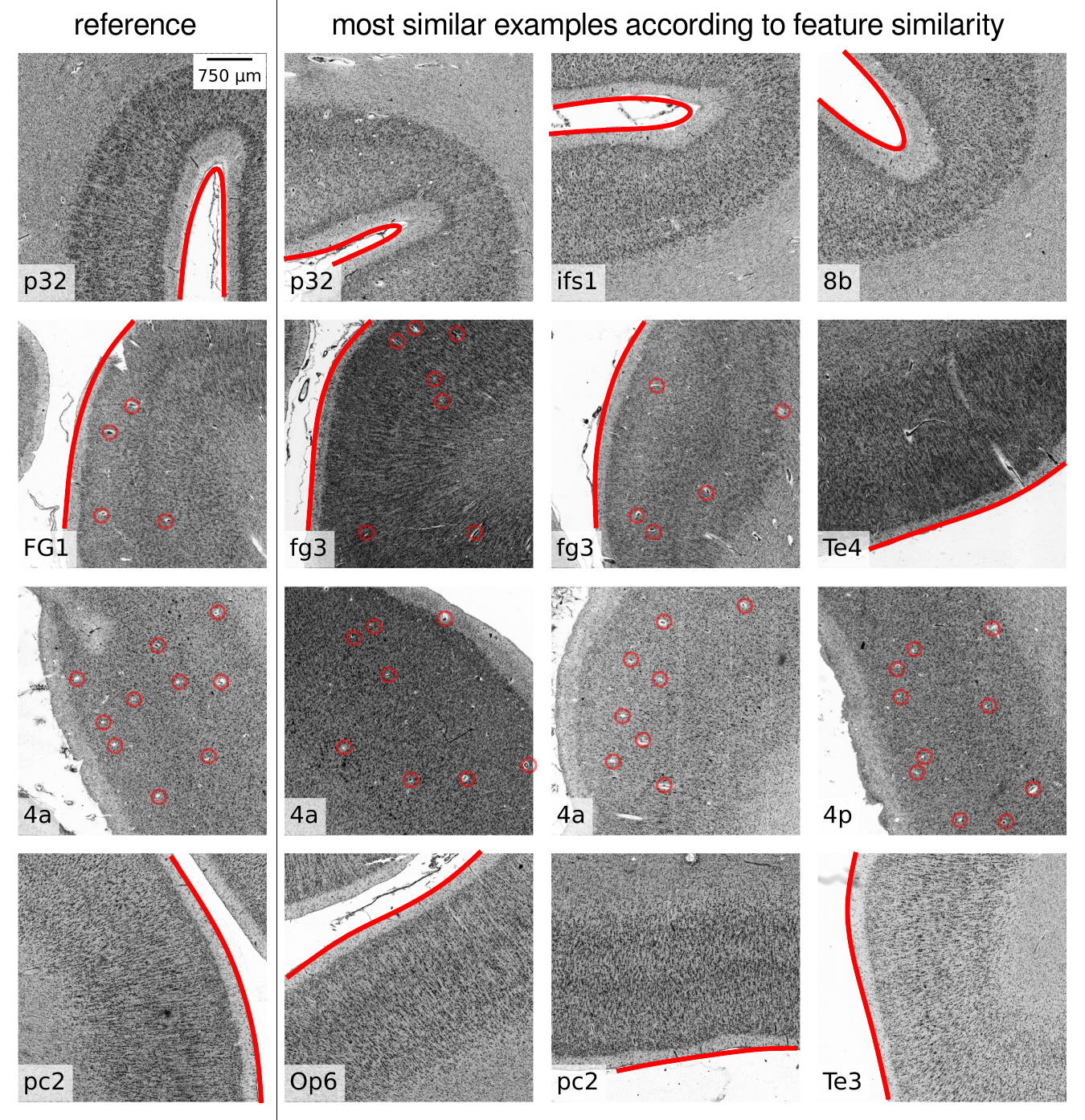}
	\end{center}
	\caption{
		\textbf{Retrieval-based analysis of features learned by SimCLR (200k), revealing shortcut learning.}
		The left column shows four randomly sampled reference image patches and their corresponding brain area.
		For each of the selected patches, the three most similar image patches from the dataset are shown, where the similarity is measured by the cosine similarity between their respective SimCLR (200k) features.
		Image similarity seems to be largely defined by tissue morphology, while being mostly independent of the brain area, and hence, cytoarchitectonic properties.
		Annotations point out possible confounding factors for the shown examples, including characteristics tissue morphology in rows 1,2,4, or characteristic blood vessel patterns in rows 2 and 3.
	}\label{fig:simclr_comparison}
\end{figure}

\cref{fig:simclr_comparison} shows a retrieval-based analysis for the SimCLR (200k) model using four randomly selected reference image patches.
Nearest-neighbor retrieval is dominated by tissue morphology and vascular patterns, while the retrieved patches often do not share the same cytoarchitectonic area.
This supports the interpretation that augmentation-based SimCLR training can learn shortcut features in this dataset, in contrast to the spatially anchored features learned by CytoNet.



\end{document}